\documentclass[5p,twocolumn,times]{elsarticle}
\usepackage{dcolumn}
\usepackage[T1]{fontenc}
\usepackage{lmodern}
\usepackage[english]{babel}
\usepackage{amsmath,amsfonts,amsthm,amssymb,amsbsy, siunitx} 
\usepackage{mathtools, cuted}
\usepackage{braket}
\usepackage{bm}
\usepackage{booktabs}
\usepackage[labelfont=bf,justification=justified ,singlelinecheck=false,font=footnotesize]{caption}
\usepackage{lineno}
\modulolinenumbers[1]
\usepackage{adjustbox}
\usepackage[para,flushleft]{threeparttable}

\journal{Journal of Molecular Spectroscopy}
\usepackage[final, authormarkuptext=name]{changes}
\definechangesauthor[name={A}, color=red]{A}
\usepackage{tablefootnote}
\usepackage{multirow}
\usepackage{url}
\usepackage{footnote}
\makesavenoteenv{tabular}
\makesavenoteenv{table}
\usepackage{amsmath}
\usepackage{txfonts}
\newcommand{\angstrom}{\text{\normalfont\AA}}

\begin{document}
	
	\begin{frontmatter}
		
		\title{Deciphering the Rotational Spectrum of the First Excited Torsional State of Propylene Oxide}   
		
		\author[kassel]{Pascal Stahl\corref{mycorrespondingauthor}}
		\ead{p.stahl@physik.uni-kassel.de}
		\author[hamburg,kiel]{Benjamin E. Arenas}
		\author[cologne]{Oliver Zingsheim}
		\author[hamburg,kiel]{Melanie Schnell}
		\author[lille]{Laurent Margulès}
		\author[lille]{Roman A. Motiyenko}
		\author[kassel]{Guido W. Fuchs}
		\author[kassel]{Thomas F. Giesen}

		\address[kassel]{Institute of Physics, University of Kassel,
              Heinrich-Plett-Str. 40, 34132 Kassel, Germany}
      \address[hamburg]{Deutsches Elektronen-Synchrotron (DESY), Notkestraße 85, 22607 Hamburg, Germany} 
    \address[kiel]{Institut für Physikalische Chemie, Christian-Albrechts-Universität zu Kiel, Max-Eyth-Str. 1, 24118 Kiel, Germany}  
     \address[cologne]{I. Physikalisches Institut, Universität zu Köln, Zülpicher Straße 77,
50937 Köln, Germany}       
      \address[lille]{Université de Lille, Laboratoire de Physique des Lasers, Atomes et Molécules (PhLAM), 2 Avenue Jean Perrin, 59650 Villeneuve-d'Ascq, France}    
        
		\cortext[mycorrespondingauthor]{Corresponding author}

\begin{abstract}
The first excited torsional state of the chiral molecule propylene oxide, $\mathrm{CH_{3}C_{2}H_{3}O}$, is investigated from millimeter up to sub-millimeter wavelengths (75-950 GHz). The first excited vibrational mode of propylene oxide, $\mathrm{\upsilon_{24}}$, is analysed using the programs ERHAM and XIAM. Rotational constants and tunneling parameters are provided, and a description of the A-E splittings due to internal rotation is given. Furthermore, the potential barrier height to internal rotation $V_{3}$ is determined to be $V_{3}=894.5079(259)\,\mathrm{cm^{-1}}$. Our results are compared with quantum chemical calculations and literature values. We present a line list of the dense spectrum of the first excited torsional state of propylene oxide in the (sub-)millimeter range. Our results will be useful for further studies of chiral molecules in vibrationally excited states, and will enable astronomers to search for rotational transitions originating from $\mathrm{\upsilon_{24}}$ of propylene oxide in interstellar space. 
\end{abstract}


\end{frontmatter}

\section{Introduction} 
Propylene oxide (PO), $\mathrm{CH_{3}C_{2}H_{3}O}$, is an oxirane (\textit{c}-$\mathrm{C_{2}H_{4}O}$) with one hydrogen being substituted by a methyl group. This molecule is one of the smallest organic chiral molecules,  
and it consists of structural elements that have been detected as separate molecules in space;
namely methane, $\mathrm{CH_{4}}$, and oxirane \citep{Fox.1978, Dickens.1997}.
In 2016, PO was detected in its vibrational ground state in the interstellar medium (ISM) towards the galactic center in Sagittarius B2(N) and was the first chiral molecule detected outside of our solar system \citep{McGuire.2016}.   
Thus, it is of high astrophysical interest.
The dense spectra of astrophysically relevant molecules can be cumbersome to analyse because line confusion can lead to difficulties in line assignment procedures. Accurate line lists help astronomers to disentangle the dense spectra ("identifying the flowers from the weeds") \citep{Herbst.2009}. Complex organic molecules (COMs), like PO,  contribute significantly to the dense spectra in the ISM, with their ground state but also with low-lying vibrationally excited states \citep{Herbst.2009}.
 COMs can not only be found in cold molecular clouds but also in the warm environment of star-forming regions, e.g. hot cores \citep{Herbst.2009}.
If PO exists in the latter warm regions, it is likely that the first vibrational state of PO is excited and should then be detectable. 
PO's isomers propanal, $\mathrm{CH_{3}CH_{2}CHO}$, and acetone, $\mathrm{CH_{3}COCH_{3}}$, were also found in space in hot core regions \citep{Snyder.2002,Friedel.2005,Lykke.2017}.
\\
\\
Propylene oxide is a small stable chiral molecule featuring hindered internal rotation, which is the large amplitude motion (LAM) of the methyl (-$\mathrm{CH_{3}}$) group with respect to the remaining frame of PO along the $\mathrm{C-CH_{3}}$ bond.
Recent laboratory investigations were triggered by the astronomical detection of PO \citep{McGuire.2016}. 
Mesko \textit{et al.} \citep{Mesko.2017} measured the sub-millimeterwave (sub-mmw) spectrum up to 1 THz and delivered a highly accurate ground state prediction using the internal rotation program XIAM \citep{Hartwig.1996}, in which the hindered internal rotation of the methyl group of PO in the vibronic ground state is  also described. The barrier height of this motion was determined to be $V_{3}=892.71(58)\,\mathrm{cm^{-1}}$. In the case of the torsional ground state, the splitting, caused by the coupling of the overall rotation and internal rotation \citep{LIN.1959,Lister.1978}, can be observed up to 400 GHz; at higher frequencies, the A and E components are blended in a single line \citep{Mesko.2017}. Mesko \textit{et al.} reported that higher torsionally excited states were seen in the spectrum, but these were not assigned due to line confusion \citep{Mesko.2017}. However, the detailed knowledge of the excited torsional states is important to further elucidate the dense spectra in the mmw and sub-mmw ranges. 
\\
\\
PO and its lowest vibrationally excited state were of high relevance for molecular spectroscopy as early as 1957, when the spectrum of PO was measured in the microwave region up to 40 GHz by Herschbach \& Swalen \citep{Swalen.1957,Herschbach.1958}.
They analysed the torsional A-E splitting of the vibrational ground state (GS), and assigned some dozens of transitions of the first excited torsional state (1$^{st}$ ETS, $\mathrm{\upsilon_{24}}=1$) and the second excited torsional state (2$^{nd}$ ETS, $\mathrm{\upsilon_{24}}=2$). In their work, they determined important structural parameters: the dipole moment components ($\mu_{a}=0.95$\,D, $\mu_{b}=1.67$\,D, and $\mu_{c}=0.56$\,D), the moments of inertia, the reduced moment of inertia of the methyl top, and the direction cosines of the internal axis with respect to the principal axes. Moreover, the threefold potential barrier height $V_{3}=895(5)~\mathrm{cm^{-1}}$ was determined \citep{Swalen.1957,Herschbach.1958}. Thus, the torsional motion of PO can be treated as a high barrier case \citep{LIN.1959,Herschbach.1958}. Isotopic investigations on $\mathrm{^{13}C}$ and D isotopologues have been carried out by Creswell \& Schwendemann \citep{Creswell.1977} and Imachi \& Kuczkowski \citep{Imachi.1983}. 
\\
\\
In this work, we present a comprehensive study of the 1$^{st}$ ETS of PO covering data from 10\,GHz to 950\,GHz, which includes the results from Herschbach \& Swalen \cite{Swalen.1957,Herschbach.1958} from 10$-$40\,GHz and newly assigned transitions from our own investigations in the 75$-$950\,GHz region.
Therefore, we used the two internal rotation data analysis programs XIAM and ERHAM \citep{Hartwig.1996, Groner.1997, Groner.2012}.
The splitting of rotational transitions of PO's 1$^{st}$ ETS was investigated by high-resolution (sub-)mmw spectroscopy. The retrieved assignment of the line splitting allowed for the determination of the potential barrier height $V_{3}$ to the respective hindered motion, which can be compared to the results of previous work on the ground state. 
These results complement the 1$^{st}$ ETS analysis from Herschbach \& Swalen and largely extend the line list with transitions from mmw and sub-mmw frequencies, where several state-of-the-art radioastronomy facilities are working, for example the Atacama Large Millimeter/submillimeter Array (ALMA), the 100\,m Greenbank Telescope (GBT), and the Effelsberg Telescope.        
\\
\\
An overview of our experiments and analysis methods is given in Section \ref{meas}, and the data analysis and results are given in Section \ref{res}. In Section \ref{disc}, the results are discussed and conclusions can be found in Section \ref{conc}.
\section{Experiment and analysis methods} \label{meas}
\subsection{Experimental setups} 
We recorded rotational spectra of PO between 75\,GHz and 950\,GHz using the BrightSpec W-band spectrometer at the DESY facility in Hamburg, the THz spectrometer in Kassel, and the Lille - Submillimeter (SMM) spectrometer. The recorded spectra cover 72\,\% of the 75$-$950\,GHz range. The availability of high-resolution laboratory data from the mmw up to the sub-mmw range allows for an accurate investigation of the rotational spectrum of PO in its 1$^{st}$ ETS including centrifugal distortion effects. 
The W-band spectrometer is a chirped-pulse Fourier-transform millimeter wave (CP-FTmmw) spectrometer, covering the frequency range from 75$-$110\,GHz. The other two spectrometers are 2$f$ modulation absorption spectrometers operating between 150\,GHz and 950\,GHz.
\begin{table*}[ht]
\centering
\caption{
Overview of the measured frequency regions included in this work, the spectrometers and their respective operating methods, the operating pressure $p$ of propylene oxide, the typical (mid-band) line width, and the frequency resolution $\Delta x_{obs}$ of the spectra. The frequency ranges are labeled by their band numbers.
}
\begin{tabular}{ccccccc}
\hline 
Band & Range /\,GHz & Instrument/Method$^{a,b}$ &  Sample &  $p$ /\,Pa & FWHM$^{c}$ /\,kHz &$\Delta x_{obs}$ /\,kHz\\
\hline

1 &75$-$110& W-band/CP-FTmmw& gas flow& 0.6&500& 60  \\
 & 75$-$110 & Kassel THz/2$f$ & static cell&8&500& 50  \\

2 &170$-$220 & Kassel THz/2$f$ &static cell&  8&500& 50  \\
&150$-$220 & Lille SMM/2$f$ &static cell&  15&300& 30\\ 

3 &260$-$330& Kassel THz/2$f$& static cell&  8&500& 50 \\
&225$-$330& Lille SMM/2$f$& static cell& 15&400& 36 \\ 

4&400$-$490& Lille SMM/2$f$& static cell& 15&600& 42 \\ 
5&490$-$660& Lille SMM/2$f$& static cell&15&800&54 \\ 
6&780$-$950 & Lille SMM/2$f$& static cell& 15& 1500 &81 \\ 
\hline 
\multicolumn{7}{l}{$^{a}$ Chirped-pulse Fourier transform millimeter wave technique} \\
\multicolumn{7}{l}{$^{b}$ 2$f$ lock-in modulation technique}\\
\multicolumn{7}{l}{$^{c}$ Full width at half maximum}\\
\end{tabular} 
\label{tab:OV}
\end{table*}
In the following, the instruments and their experimental details are briefly introduced, and the information is summarised in Table \ref{tab:OV}. The frequency resolution $\Delta x_{obs}$ for measured line frequencies are between 36\,kHz and 81\,kHz depending on the used experimental setup and the frequency range.  
\\
\\
The W-band spectrometer is described in detail in Arenas \textit{et al.} \citep{Arenas.2017}; the gas-phase molecules were excited by 500 ns long excitation pulses, which were produced by an arbitrary waveform generator and frequency-upconverted into the mmw region.  The PO spectrum was obtained by co-adding 500,000 free induction decays, which were recorded for 4\,$\muup$s each.  This was performed in the segmented fashion \citep{Neill.2013}, whereby the 35\,GHz bandwidth spectrum was measured in 30\,MHz segments. The resulting total measurement time was about 50 minutes. 
The liquid PO sample was placed in a reservoir external to the vacuum chamber, treated in a freeze-pump-thaw manner, and probed with the chamber acting as a room-temperature  slow flow gas cell. The sample pressure in the chamber was maintained at approximately 0.6\,Pa.   
\\
\\
The Kassel THz spectrometer was used, utilizing a 2$f$ frequency lock-in modulation technique, to probe the sample in a static vacuum glass cell with a total length of 3\,m. The liquid sample was stored in a long-necked flat-bottomed glass flask at room temperature and connected to the glass cell via a needle valve. The operating pressure was 5\,Pa. The high vapour pressure of PO at room temperature of roughly 580\,hPa \citep{Bott.1966} is sufficient to fill the cell without heating the sample. 
The spectra were taken in the regions 170$-$220\,GHz and 260$-$330\,GHz with a frequency resolution of $\Delta x_{obs}=50$\,kHz.
The spectrometer's details and data reduction description are summarised in more detail in Herberth \textit{et al.} \citep{Herberth.2019} and Stahl \textit{et al.} \citep{Stahl.2020}. 
\\
\\
Absorption measurements of PO using the fast-scan terahertz spectrometer in Lille were performed between 150\,GHz and 950\,GHz. The details of the Lille - Submillimeter (SMM) spectrometer are described in Zou \textit{et al.} \citep{Zou.2020} and Motiyenko \textit{et al.} \citep{R.A.Motiyenko.2019}. 
A commercially available VDI frequency multiplication chain driven by a home-made fast sweep frequency synthesiser was used as the radiation source. The fast sweep system is based on the up-conversion of an AD9915 direct digital synthesiser (DDS) operating between 320 and 420\,MHz into the Ku band (12.5–18.25\,GHz) by mixing the signals from the AD9915 and an Agilent E8257 synthesiser with subsequent sideband filtering. The DDS provides rapid frequency scanning with up to 50\,$\muup$s per point frequency switching rate. In order to improve the signal-to-noise ratio (SNR), the spectrum was scanned with a slower rate of 1\,ms per point and with 4 scans co-averaged. The sample pressure during measurements was about 15\,Pa at room temperature. Absorption signals were detected using an InSb liquid He-cooled bolometer (QMC Instruments Ltd.).

\subsection{Quantum chemical calculations}
We performed quantum chemical anharmonic frequency calculations based on geometry optimisations at the B3LYP/aug-cc-pVTZ, MP2/def2-TZVP, and MP2/6-311G(d,p) levels of theory using Gaussian 16 \citep{Frisch.2016}. The rotational constants of the 1$^{st}$ ETS were calculated. Additionally, the full set of centrifugal distortion constants up to the sextic order of the GS was retrieved and later used as the  
parameter set for the 1$^{st}$ ETS at the beginning of the assignment procedure.
Furthermore, we calculated the barrier height to internal rotation $V_{3}$ (harmonic calculation), which will be discussed in Section \ref{res}. The results of our calculations are given in Table \ref{tab:abini} and will be discussed below. 
\begin{table*}[ht]
\centering
\caption{Initial molecular parameters in the Watson A-reduction for the first excited torsional state (1$^{st}$ ETS) of PO, $\mathrm{\upsilon_{24}}=1$, based on quantum chemical anharmonic frequency calculations. The centrifugal distortion parameters are those calculated for the GS. Harmonic and anharmonic vibrational energies of the $\mathrm{\upsilon_{24}}=1$ state are given. The calculated electric dipole moment components $\mu_{a}$, $\mu_{b}$, $\mu_{c}$, and the total dipole moment $\mu_{tot}$ of PO's equilibrium structure are given in Debye.
}
\begin{tabular}{llccccc}
\hline 
&\multirow{2}{*}{Parameter}&  B3LYP	&  MP2 & MP2 \\
&  &	 aug-cc-pVTZ & def2-TZVP & 6-311G(d,p) \\
  \hline
 \multirow{5}{*}{1$^{st}$ ETS}&$E_{harm}$ /$\mathrm{cm^{-1}}$ & \tablenum{ 207.05 }& \tablenum{ 217.01}  & \tablenum{ 214.47}\\  
&$E_{anharm}$ /$\mathrm{cm^{-1}}$ & \tablenum{  196.47} & \tablenum{ 205.88 }  & \tablenum{ 204.12} \\   
&$A$ /MHz & \tablenum{ 18056.20} & \tablenum{  17896.23} &  \tablenum{17875.25}\\ 
&$B$ /MHz & \tablenum{ 6586.44 } & \tablenum{6655.33} &  \tablenum{6624.06} \\ 
&$C$ /MHz & \tablenum{ 5870.06 }  & \tablenum{5924.47 }&  \tablenum{5905.01}\\ 
\hline
 \multirow{9}{*}{GS}&$\Delta_{J}$ /kHz   & \tablenum{2.88}   &  \tablenum{2.91} &  \tablenum{2.84}  \\ 
&$\Delta_{JK}$ /kHz  &  \tablenum{3.59}&  \tablenum{3.37 }&  \tablenum{3.24} \\ 
&$\Delta_{K}$ /kHz  &   \tablenum{19.87 }&  \tablenum{19.17 } &  \tablenum{18.93 }  \\ 
&$\delta_{J}$ /kHz   & \tablenum{ 0.19 }& \tablenum{0.19} &  \tablenum{0.18} \\ 
&$\delta_{K}$ /kHz  &  \tablenum{2.60} &  \tablenum{2.63} & \tablenum{ 2.59} \\ 
&$\Phi_{J}$ /Hz      &  \tablenum{ 0.0017} &   \tablenum{   0.0017} & \tablenum{  0.0016}\\ 
&$\Phi_{JK}$ /Hz   &  \tablenum{-0.0066 }&  \tablenum{-0.0050 }& \tablenum{ -0.0040} \\ 
&$\Phi_{KJ}$ /Hz    &  \tablenum{ 0.028 } &  \tablenum{ 0.022} & \tablenum{ 0.018 }\\   
&$\Phi_{K}$ /Hz  & \tablenum{ 0.044} & \tablenum{  0.042 }& \tablenum{ 0.046 } \\ 
\hline
 \multirow{4}{*}{equilibrium}&$\mu_{a}$ /D  & \tablenum{ 0.88} & \tablenum{ 0.92 }& \tablenum{ 0.87 } \\ 
&$\mu_{b}$ /D  & \tablenum{ 1.67} & \tablenum{ 1.70 }& \tablenum{ 1.67 } \\ 
&$\mu_{c}$ /D  & \tablenum{ 0.67} & \tablenum{  0.48 }& \tablenum{ 0.48 } \\ 
&$\mu_{tot}$ /D  & \tablenum{ 2.00} & \tablenum{  1.95 }& \tablenum{ 1.99 } \\ 
\hline
\end{tabular} 
\label{tab:abini}
\end{table*}

\subsection{Analysis Methods} \label{AM}
\begin{figure}[ht]
\centering
\includegraphics[width=\hsize]{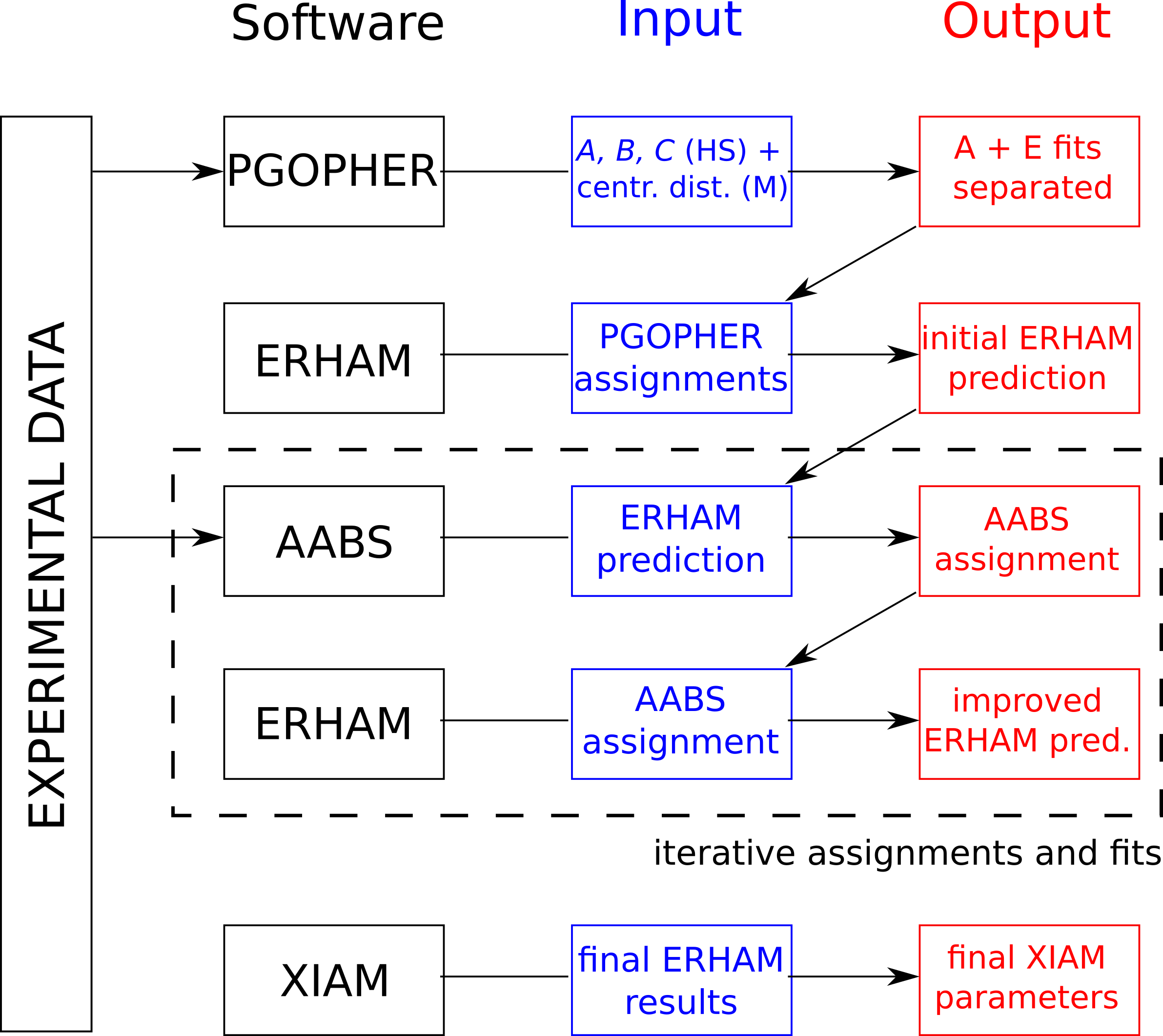}
\caption{
Schematic diagram of the analysis methods procedure for the assignment of the first torsional excited state of propylene oxide. The PGOPHER input $A$, $B$, and $C$ are the rotational constants from Herschbach \& Swalen \citep{Herschbach.1958} and the centrifugal distortion constants are from Mesko \textit{et al.} \citep{Mesko.2017}.
}
\label{fig:schem}
\end{figure}

At room temperature, PO has a dense spectrum, where rotational transitions from the GS as well as from the ETSs can be readily assigned.
The transitions from the vibronic GS are well known \citep{Mesko.2017}, thus, most of the remaining strong features in the spectra can be assumed to belong to vibrationally excited states. This residual spectrum is used as a basis for further analysis of the 1$^{st}$ ETS. 
The 1$^{st}$ ETS is the lowest vibrationally excited state, lying about 200\,$\mathrm{cm^{-1}}$ above the GS (see Table \ref{tab:abini}) and is thereby the most intense vibrationally excited state.
Lines of the 1$^{st}$ ETS have around 38\% of the intensity of comparable GS transitions\footnote{Assuming a Boltzmann intensity distribution at room temperature and neglecting higher vibrationally excited states.} at room temperature.
As the PO spectra have a high line density and also show internal molecular motion, the line assignment is elaborate. The analysis was done in several steps using different kinds of analysis software, which is schematically shown in Figure \ref{fig:schem}. 
At the beginning, we used the PGOPHER software \citep{Western.2014, Western.2017} and the experimental data from Kassel for first assignments by employing the experimentally derived rotational constants $A$, $B$, and $C$ from Herschbach and Swalen \citep{Swalen.1957,Herschbach.1958} for the A and E states and the centrifugal distortion constants of the GS  \citep{Mesko.2017}. The A state transitions of an asymmetric molecule with internal rotation can be described similarly to an asymmetric rigid-rotor molecule \citep{Swalen.1957}, and thus we first assigned the strongest A state lines. We then applied this procedure for the assignment of strong E state transitions that were within about 10$-$20\,MHz of the respective A state transitions. 
PGOPHER only allows for a description of weak internal rotation effects\footnote{See PGOPHER release notes: \url{http://pgopher.chm.bris.ac.uk/download/old/9.0.101/Help/bugs.htm}, last accessed 17th, September 2020.} or of the description of the A and E states separately (like two individual molecules). 
\\
\\
Thus, we moved on to the internal rotation software ERHAM \citep{Groner.1997}, which delivered an initial ERHAM prediction based on the PGOPHER assignments (see Figure \ref{fig:schem}). The already known low-frequency assignments from Herschbach and Swalen \citep{Swalen.1957,Herschbach.1958} have been included in this analysis.
Due to the ability to label and assign the A and E states of PO in one combined line list, we employed the AABS package \citep{Kisiel.2005, Kisiel.2012} available from the PROSPE website  \citep{Demaison.2001}.
Iteratively, the ERHAM prediction, based on the ERHAM least-squares fit analysis, was used to aid the assignment of the experimental data from Band 1 to Band 6 from DESY, Kassel, and Lille in the JPL/CDMS (SPFIT/SPCAT)\footnote{Although we used the JPL/CDMS (SPFIT/SPCAT) format, we did not use the spectral analysis software SPFIT/SPCAT.} data format \citep{PICKETT.1998,Muller.2005}.
The AABS assignment was used as the ERHAM input, which delivered an improved ERHAM prediction after the fitting process. We started with low $J$, $K_{a}$ transitions ($J$, $K_{a}$<10, 5), rotational parameters up to the quartic order, and with just a few ERHAM structure and tunneling parameters, i.e., $\rho$, $\beta$, $\epsilon_{1}$, and $[A-(B+C)/2]_{1}$ (these parameters will be discussed later). We gradually included higher $J$ and $K$ transitions at higher frequencies, which led us to introduce additional centrifugal distortion and tunneling parameters.
These iterations proceeded until a reasonable ERHAM fit was achieved (Tables \ref{tab:resultsI} and  \ref{tab:resultsII}). The highest quantum numbers included in our assignment are $J'',K_{a}'',K_{c}''=54,24,54$ for the lower state. Since ERHAM uses the quantum number format $J, N$ with $N=K_{a}-K_{c}+J+1$\footnote{$N$ was taken from the LINERH.for code available from the PROSPE website \citep{Demaison.2001}.}, we used a self-written Python program to convert the quantum labels. Almost exclusively, transitions with an intensity $I$ (at 300\,K, $J_{max}=82$, partition function $Q=295409.393$) stronger than the decadic logarithmic intensity $I$=$-4.0\,\mathrm{nm^{2}MHz}$ in the JPL/CDMS format \citep{PICKETT.1998,Muller.2005} were assigned, except for the sensitive W-band measurement, where also weaker ($I$>$-$6.0$\,\mathrm{nm^{2}MHz}$) transitions were included. Q-branch transitions were assigned up to 460\,GHz. At higher frequencies, the Q-branch intensities decreased and mostly strong R-branches were observed. ERHAM \citep{Groner.1997} is not able to directly determine the potential barrier height\footnote{Note that Peter Groner's program BARRIER is able to calculate $V_{3}$ from the splittings from the ERHAM analysis.}, and thus, we decided to use the internal rotation program XIAM as an additional analysis software, which also provides a useful comparison between the two programs.
\\
\\
We used the ERHAM results and almost\footnote{In the final XIAM fit, 19 more transitions have been included compared to ERHAM because a few assignments that resulted in large residuals in the final ERHAM and XIAM fit were omitted. No conspiciuous correlations of these assignments with respect to the quantum numbers were found.} the same line assignments in our XIAM analysis.
However, there are a few constraints that have to be considered when going from one program to the other: XIAM is based on the \textit{internal axis method} \citep{Hartwig.1996,Kleiner.2010}, and
ERHAM uses the "Effective Rotational HAmiltonian Method", where the internal rotor Hamiltonian is set up in the $\rho-$axis system and later transformed to the principal axis system \citep{Groner.2012}. Thus, both programs have different coordinate systems\footnote{The angles of the XIAM analysis are denoted with the subscript \textit{X} as can be seen in Table \ref{tab:resultsII}.} and different tunneling parameters. 
Furthermore, XIAM uses the JPL/CDMS quantum number format $J, K_{a}, K_{c}$ \citep{PICKETT.1998,Muller.2005}.
Thus, the different quantum labels and data format of XIAM and ERHAM were interconverted using a self-written Python program. 
The results from our XIAM analysis, including the barrier height to internal rotation $V_{3}$, are given in Tables \ref{tab:resultsI} and \ref{tab:resultsII}.

\section{Spectral analysis and results} \label{res}
\subsection{Spectral analysis}
\begin{figure*}[ht]
\centering
\includegraphics[width=\hsize]{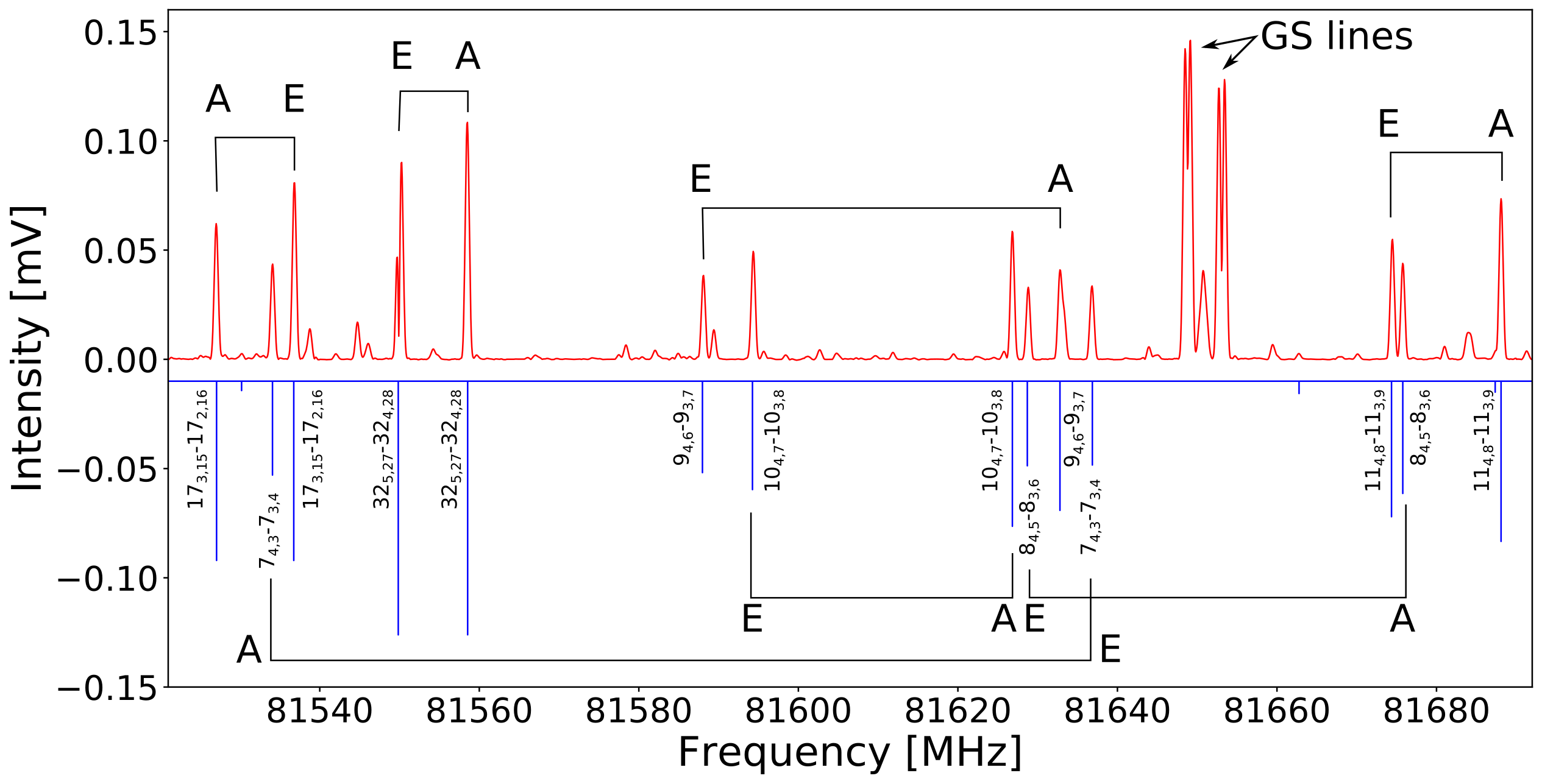}
\caption{
Spectrum of propylene oxide around 81.6\,GHz (upper trace, red) and a room temperature ERHAM prediction (stick spectrum, lower trace, blue), based on a least-squares-fit analysis (Tables \ref{tab:resultsI} and  \ref{tab:resultsII}), with quantum number labels and A and E state designation. Ground state (GS) transitions around 81.65\,GHz can be also seen. The simulated trace was given an offset for improved visibility. 
}
\label{fig:100}
\end{figure*}

\begin{figure*}[ht]
\centering
\includegraphics[width=\hsize]{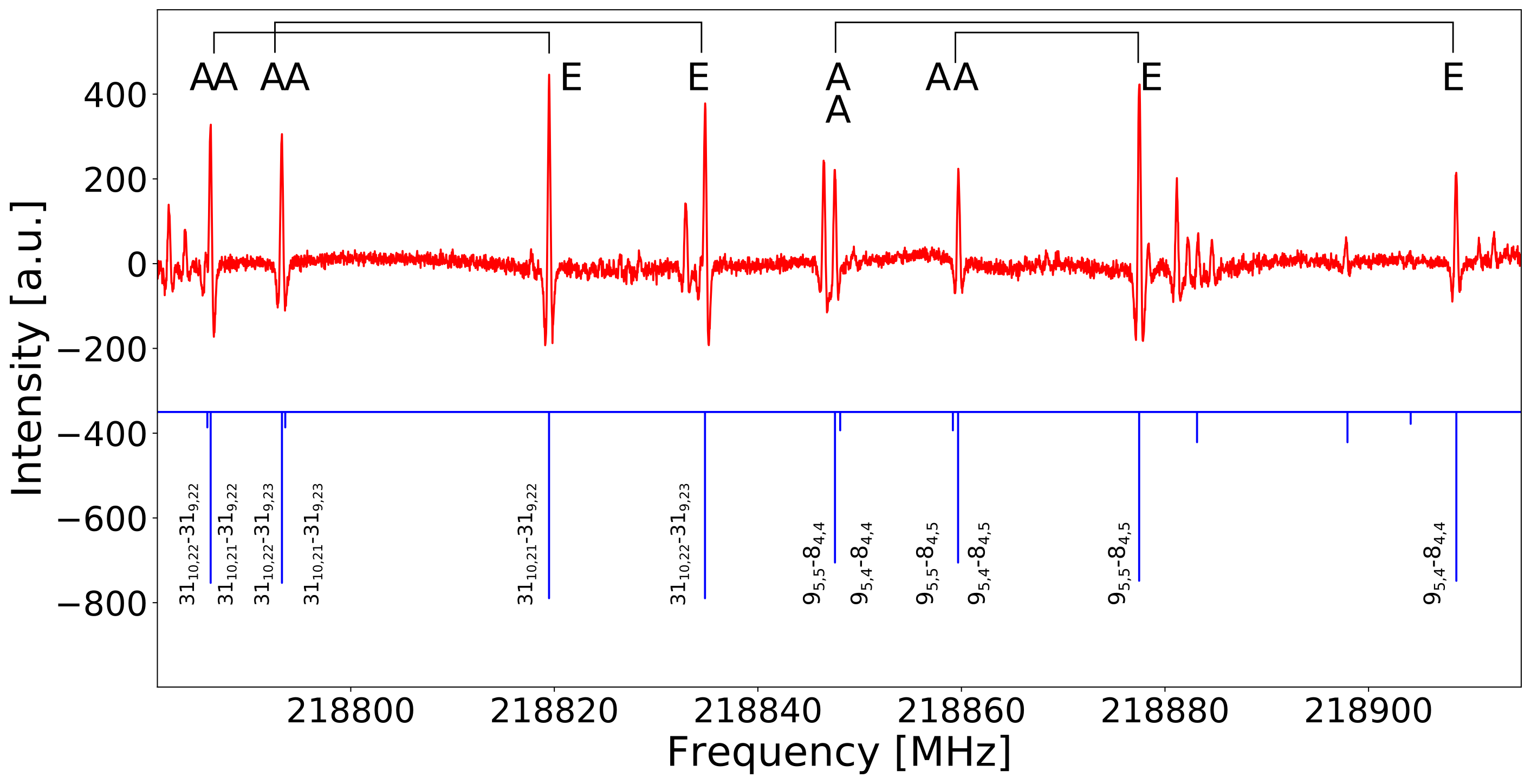}
\caption{
Spectum of propylene oxide around 218.8 GHz (upper trace, red) and a room temperature ERHAM prediction (stick spectrum, lower trace, blue), based on a least-squares-fit analysis (Tables \ref{tab:resultsI} and \ref{tab:resultsII}), with quantum number labels and A and E state designation. The simulated trace was given an offset for improved visibility.
}
\label{fig:200}
\end{figure*}

\begin{figure*}[t]
\centering
\includegraphics[width=\hsize]{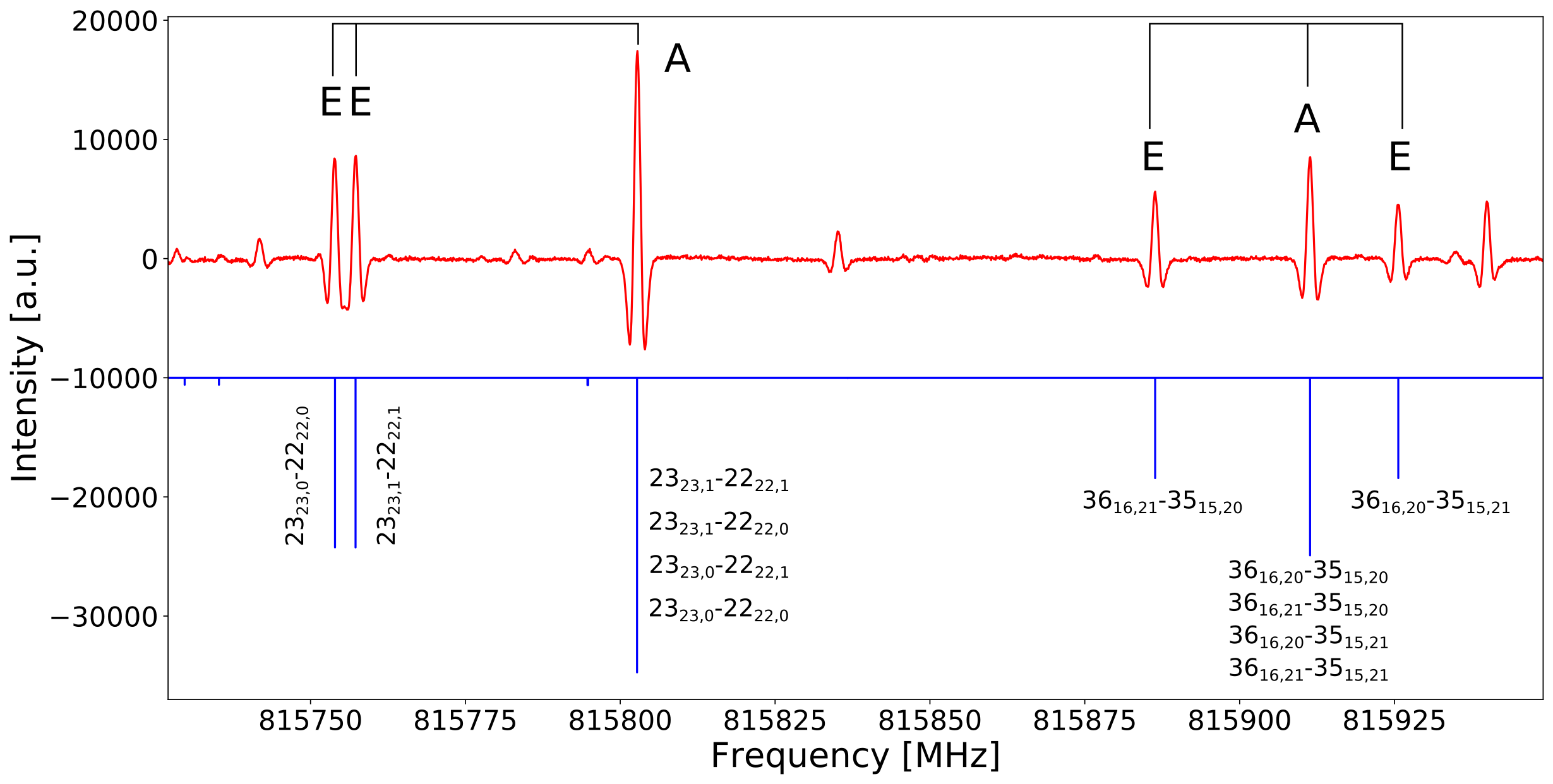}
\caption{
Spectrum of propylene oxide around 815.8\,GHz (upper trace, red) and a room temperature ERHAM prediction (stick spectrum, lower trace, blue), based on a least-squares-fit analysis (Tables \ref{tab:resultsI} and \ref{tab:resultsII}), with quantum number labels and A and E state designation. The R-branch transitions show the expected triplet pattern of the $\mathrm{AA}$, $\mathrm{E}$, and $\mathrm{E}$ levels with an intensity ratio of 2:1:1. The simulated trace was given an offset for improved visibility.
}
\label{fig:800}
\end{figure*}

The spectrum of the 1$^{st}$ ETS shows patterns similar to those observed in the ground state spectrum and consists of strong regularly emerging R-branch and Q-branch transitions, the latter ones are separated by about $2\cdot(A-\dfrac{B+C}{2}) \approx 23.4~\mathrm{GHz}$. 
In Figures \ref{fig:100}, \ref{fig:200}, and \ref{fig:800}, spectral excerpts of our measurements around 80\,GHz, 218\,GHz, and 816\,GHz are shown. 
The lines mostly emerge as doublets or triplets. In the case of a triplet, three separate lines (blended AA, E, and E) can be observed, where the two asymmetry sides ($J=K_{a}+K_{c}$ and $J+1=K_{a}+K_{c}$) are blended for the A components, but separated for the E ones \citep{Herschbach.1958}.
In the case of a doublet, there is a splitting into two lines, AA and EE, this time both E components are also blended.
In contrast to the ground state splittings in PO, which are  typically on the order of 0.1$-$1\,MHz (see strong lines around 81.65\,GHz in Figure \ref{fig:100}), the size of splitting of the ETS lines is much larger and is up to hundreds of\,MHz; a fact that makes the line assignment of ETS transitions more difficult than those of the GS. In the case of the 1$^{st}$ ETS, most A-E splittings are between 10 and 50 MHz and could be observed throughout the whole frequency range up to 1\,THz. Even larger splittings up to more than 100\,MHz are also observed and assigned.
In dense spectra, for example in the case of the PO spectrum in the region 104.66$-$104.76\,GHz with more than 30 measured transitions within 100\,MHz, an unambiguous assignment of lines split by 50\,MHz to 100\,MHz is challenging.     
As an example, Figure \ref{fig:100} shows several Q-branch ($J'=J''$) b-type transitions with the notation $J'_{K_{a},K_{c}}\leftarrow J''_{K_{a}-1,K_{c}+1}$ with pairs of A and E states.
The seven A-E pairs highlighted in this spectral excerpt exhibit different magnitudes of splitting. For example, the $7_{4,3}\leftarrow 7_{3,4}$ transition has an A-E splitting of 102\,MHz. 
As expected from spin statistics \citep{Lister.1978}, in most cases, the A and E state doublets have roughly the same intensity (ratio 1:1), which can be seen in Figure \ref{fig:100}. The typical SNR for stronger lines in the W-band is about 600:1. The line width, the full width at half maximum (FWHM), varies between 500 and 600 kHz. 
Figure \ref{fig:200} shows an excerpt of the recorded PO spectrum around 218 GHz with $\mathrm{Q_{9}}$-branch ($J''=31$) and $\mathrm{R_{4}}$-branch ($J''=8$) transitions. Both exhibit A and E level transitions as labeled with their respective quantum numbers. The typical SNR in this region is about 180:1, and the FWHM varies from 500\,kHz (in the case of no blending) to 800\,kHz (in the case of small AA or AE splitting). 
In Figure \ref{fig:800}, two triplets belonging to the 1$^{st}$ ETS are shown with their respective quantum numbers. Both triplets originate from R-branches and show the characteristic intensity ratio of 2:1:1 (AA:E:E). Furthermore, weak transitions from higher vibrationally excited states can be seen. The A states with $J'',K''_{a}=22,22$ and $J'',K''_{a}=35,15$ are blended, the corresponding E states are split by 3 MHz and 40 MHz.

\subsection{Spectroscopic results} \label{specres}
\begin{figure}[t]
\centering
\includegraphics[width=\hsize]{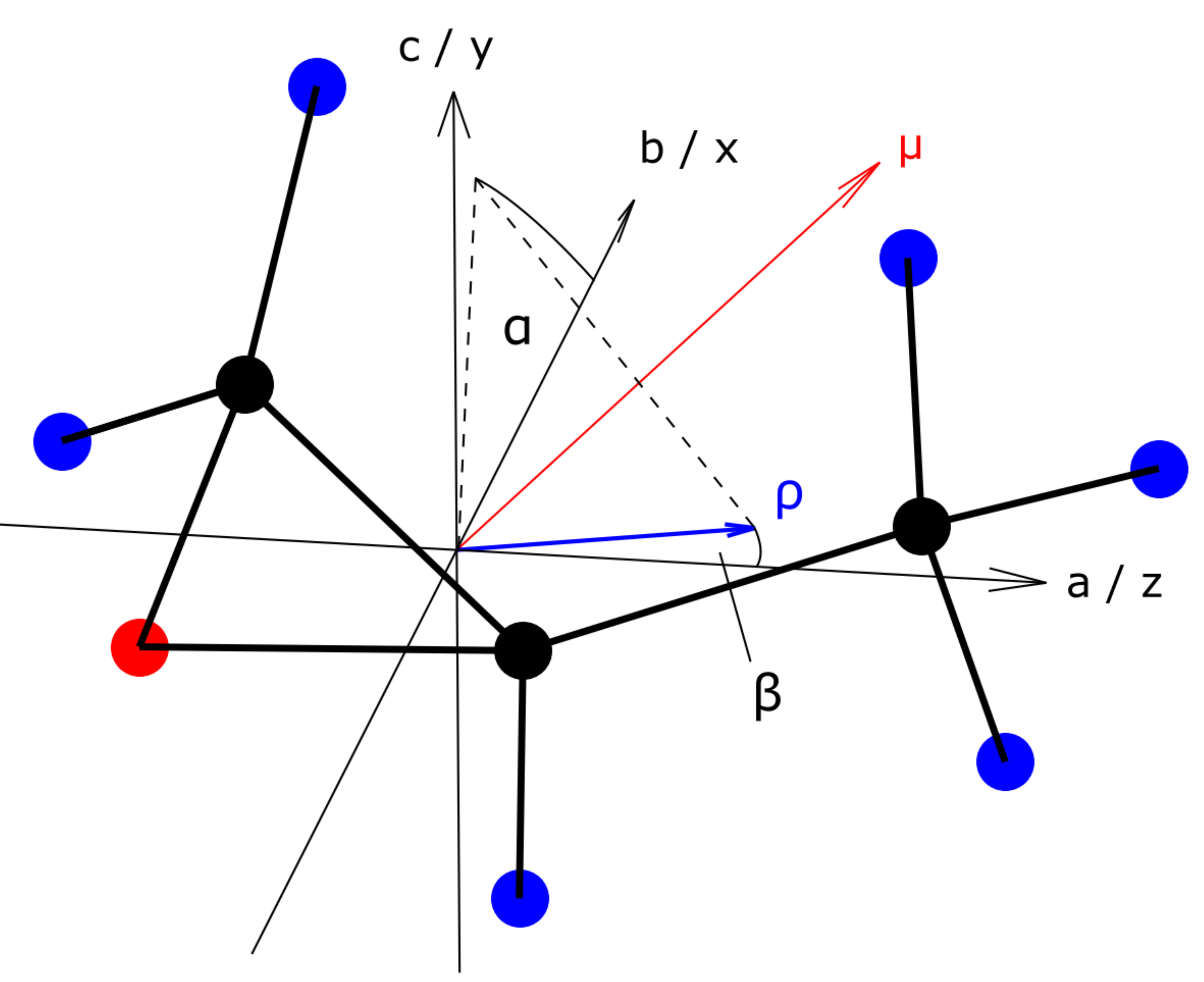}
\caption{
Structure of propylene oxide in the principal axis system (a, b, c) and the corresponding (x, y, z)-system in $\mathrm{I_{r}}$ representation. The atomic coordinates (full circles) are based on B3LYP/aug-cc-pVTZ calculations using Gaussian 16. Hydrogen atoms are depicted in blue, carbon in black, and oxygen in red. The $\rho$-axis vector and the angles $\beta~\left[ \angle(\textcolor{blue}{\rho},a)\right] $ and $\alpha~\left[ \angle(\textcolor{blue}{\rho}-\mathrm{projection~onto}~b-c-\mathrm{plane},a-b-\mathrm{plane})\right] $ are taken from the ERHAM results (Table \ref{tab:resultsII}). 
The dipole moment vector $\mu$ is given in red.
}
\label{fig:ERHAM_structure}
\end{figure}

The results for our data analysis of the 1$^{st}$ ETS of PO in the vibrationally excited state $\mathrm{\upsilon_{24}}$ using ERHAM and XIAM are summarised in Tables \ref{tab:resultsI} and \ref{tab:resultsII} together with those from Herschbach and Swalen \citep{Herschbach.1958}.
\begin{table*}[ht]
\centering
\caption{Rotational constants, quartic and sextic centrifugal distortion constants, and tunneling parameters for the first excited torsional  state $\mathrm{\upsilon_{24}}$ of propylene oxide, with uncertainties (given in brackets for the least significant digits), described in the Watson A-reduction. Data from Herschbach \& Swalen \citep{Herschbach.1958} are given for comparison.
}
\begin{tabular}{lccccc}
\hline 
\multirow{2}{*}{Rotational parameter}& \multicolumn{2}{c}{This work}  & \multicolumn{3}{c}{Herschbach \& Swalen} \\
  & ERHAM& XIAM	& A state & E state & (A+2E)/3 \\
  \hline

$A$ /MHz & 18014.521213(327)  &18014.526013(613)& 18010.82$^{a}$ &18016.15$^{a}$ & 18014.37  \\ 
$B$ /MHz & 6670.114709(91) & 6670.816226(177)& 6669.93$^{a}$ & 6669.96$^{a}$ & 6669.95\\ 
$C$ /MHz & 5944.996211(98) & 5944.296854(191)&5944.82$^{a}$ & 5945.05$^{a}$& 5944.97\\ 
$\Delta_{J}$ /kHz &2.8998047(745) & 2.899985(139) &\multicolumn{2}{c}{2.97(50)$^{b}$}   \\ 
$\Delta_{JK}$ /kHz &3.535792(527) & 3.614558(1037)  & \multicolumn{2}{c}{4.2(20)$^{b}$}  \\ 
$\Delta_{K}$ /kHz &19.51430(115) & 19.455234(1524)  & \multicolumn{2}{c}{19.50(50)$^{b}$} & \\ 

$\delta_{J}$ /kHz& 0.18809133(105)  & 0.188434(21) &    &  \\ 
$\delta_{K}$ /kHz &  2.179023(556)  & 1.338458(1061)  & &  \\ 

$\Phi_{J}$ /Hz & 0.0015221(176) &  0.001492(33)  &  &  &     \\ 
$\Phi_{JK}$ /Hz &  -0.014163(216) &  -0.011949(399)   &  & &   \\ 
$\Phi_{KJ}$ /Hz & 0.051390(728) & 0.089568(1420)      & & &  \\ 
$\Phi_{K}$ /Hz & 0.02125(106)&     &    & &  \\ 
\hline

$\epsilon_{1}$ /MHz   &76.87272(788)  &  & &  & \\
$[A-(B+C)/2]_{1}$ /kHz   & -5.1564(793)   &  & &  & \\
$[(B+C)/2]_{1}$  /kHz  &  0.7135(233) &  & &  & \\
$[\Delta_{JK}]_{1}$  /Hz  & 1.3171(790) &  & &  & \\

$[\delta_{K}]_{1}$  /Hz  &  0.963(112)   &  & &  & \\
$[\Phi_{J}]_{1}$   /Hz  &  -0.00002487(269) &  & &  & \\
\hline
$\Delta_{\pi2J}$ /MHz   &    & 0.141435(2531) & &  & \\
$\Delta_{\pi2K}$ /MHz   &    & -0.174080(9906)  & &  & \\
$\Delta_{\pi2-}$ /MHz    &    & -0.0336(37)  & &  & \\
\hline
$\kappa$  &  -0.87984  &  -0.87962 & &  &  \\
$\sigma$ /kHz & 87.3 & 182.9\\ 
$\sigma_{err}$ & 1.67 & 3.40\\
No. of lines (unblended)& 4237  & 4980 && \\ 
No. of lines (blended)& 4961  & 4980&&\\ 
\hline 
\multicolumn{6}{l}{$^{a}$ These parameters are taken from the 1$^{st}$ ETS analysis of Herschbach \& Swalen \citep{Herschbach.1958}.}\\
\multicolumn{6}{l}{~~~No uncertainties were given for the 1$^{st}$ ETS rotational  constants.}\\
\multicolumn{6}{l}{$^{b}$ These parameters are taken from the ground state analysis of Herschbach \& Swalen \citep{Herschbach.1958}.}
\end{tabular} 
\label{tab:resultsI}
\end{table*}
\begin{table*}[t]
\centering
\caption{Structural parameters and barrier height for the first excited torsional state $\mathrm{\upsilon_{24}}$ of propylene oxide, with uncertainties (given in brackets for the least significant digits), described in the Watson A-reduction. The data from Herschbach \& Swalen \citep{Herschbach.1958} are given for comparison.
}
\begin{tabular}{lccc}
\hline
Structural & \multicolumn{2}{c}{This work}  & Herschbach \& Swalen \\
 parameter & ERHAM& XIAM	& A \& E state  \\
  \hline
$\rho$    &  0.10252334(449) &  0.102880781$^{a}$ &  0.103$^{b}$  \\
$\beta$  /$^\circ$  &  9.54061(345)  &      &9.41942$^{b}$  \\
$\alpha$ /$^\circ$    &   88.280(373)    &  &   \\
$\gamma_{X}$ /rad&  & -1.616515110$^{b}$ \\
$\beta_{X}$ /rad&  & 0.166484989$^{b}$ \\
$\epsilon_{X}$ /rad&  &  -1.61154(644)  \\
$\delta_{X}$ /rad&  & 0.47094(95)  \\
$I_{\alpha}$ /$\mathrm{u\angstrom^{2}}$    &  3.183008(265) &  3.193996693$^{b}$& 3.194$^{a}$  \\
$I_{r}$ /$\mathrm{u\angstrom^{2}}$    &  2.871687(236)&  &  \\
$\angle(i,a)$ /$^\circ$    &   26.98711(770)&  26.9828(54) & 26.86$^{b,c}$  \\
$\angle(i,b)$ /$^\circ$    &  89.304(151) &  91.059(168) &89.74$^{b,c}$  \\
$\angle(i,c)$ /$^\circ$    &  63.02333(431)&  63.0415(125)&63.14$^{b,c}$ \\
$F_{0}$ /GHz    &  158.7741$^{d}$ &  158.2278$^{e}$ &  \\
$F$ /GHz   &  175.9868(144) &  175.447889628$^{b}$ &   \\
$F$ /$\mathrm{cm^{-1}}$    &  5.870288(482) &  5.85231166 & 5.856  \\
$V_{3}$ /$\mathrm{cm^{-1}}$  &  898.154$^{f}$ &   894.5079(259) & 895(5) \\
$s$  &   &   67.931979$^{a}$ &68.0 \\
$\Delta_{0}$ /MHz$^{g}$    &  -230.6182(236)&  &-227.9000  \\
\hline 
\multicolumn{4}{l}{$^{a}$ These values were derived from the XIAM program during the analysis from the fit structural }\\
\multicolumn{4}{l}{~~ parameters and no uncertainties were given from XIAM.}\\
\multicolumn{4}{l}{$^{b}$ These parameters are taken from the ground state analysis of Herschbach \& Swalen  \citep{Herschbach.1958}.} \\
\multicolumn{4}{l}{$^{c}$ Calculated from the direction cosines $\lambda$ in Herschbach \& Swalen \citep{Herschbach.1958}.}\\
\multicolumn{4}{l}{$^{d}$ Calculated from $I_{\alpha}$ using Equation (\ref{form:F0}).}\\
\multicolumn{4}{l}{$^{e}$ Fixed value from Herschbach \& Swalen \citep{Herschbach.1958}.} \\
\multicolumn{4}{l}{$^{f}$ The potential barrier height in the ERHAM analysis was determined with Equation (\ref{barrier}) using the }\\
\multicolumn{4}{l}{~~ ERHAM value for $F$ and the parameter $s=68.0$ from Herschbach \& Swalen \citep{Herschbach.1958}}.\\
\multicolumn{4}{l}{$^{g}$ $\Delta_{0}=E_{A}-E_{E}$ torsional energy difference of the energy levels of the A and E state, respectively.} \\
\end{tabular} 
\label{tab:resultsII}
\end{table*}
In our Hamiltonian, we used the primary constants $A$, $B$, and $C$, the quartic distortion parameters $\Delta_{J}$, $\Delta_{JK}$, $\Delta_{K}$, $\delta_{J}$ and $\delta_{K}$, as well as the sextic distortion constants $\Phi_{J}$, ..., $\Phi_{K}$. Furthermore, we fit the tunneling energy parameter $\epsilon_{1}$ and five tunneling parameters related to rotational constants and centrifugal distortion constants, which are denoted by  square brackets, i.e., $\mathrm{[~~]_{q}}$ \citep{Groner.2012}. For PO, there is only one LAM originating from the methyl group ($-\mathrm{CH_{3}}$). Thus, q'=0 and will be omitted from the notation $\mathrm{[~~]_{q}}$. Furthermore, there is just one set of the $\rho$-axis vector and the angles $\beta$ and $\alpha$, and the subscript "1" is omitted. For further information on the notation in ERHAM, see Ref. \citep{Groner.2012}.
\\
At the beginning of the ERHAM fitting process, the structure parameters $\rho$, $\beta$, and $\alpha$\footnote{These additional structural parameters are related to the direction cosines $\lambda_{i}$ between the internal axis $i$ and the principal axes $a$, $b$, and $c$.} 
varied when including new transitions to the fit. We therefore fixed the $\rho$ value to that reported by Mesko \textit{et al.} \citep{Mesko.2017}. The values of the angles $\beta$ and $\alpha$ were deduced from the optimised geometry and basis transformation matrix of the B3LYP/aug-cc-pVTZ prediction. During the assignment of the first hundreds of lines up to 300 GHz, the structure and tunneling parameters did not converge to reliable values. The fit converged to robust results only when adding the high frequency transitions and fitting $\beta$ and $\alpha$. 
The final ERHAM results with the torsional structural parameters $\rho$, $\beta$, and $\alpha$ allowed for the precise determination of the structure of PO and its internally rotating methyl group, which is shown in Figure \ref{fig:ERHAM_structure}. The atomic coordinates and the dipole moment vector are based on B3LYP/aug-cc-pVTZ calculations. The $\rho$-axis vector is strongly aligned along the principal axis a, which can be seen from the $\rho$-axis vector components $\rho_a=0.10110529$, $\rho_b=0.00050994$, and $\rho_c=0.01698524$, which were calculated from $\rho$, and the angles $\beta$ and $\alpha$ by a spherical coordinate transformation using a self-written python code. Thus, the $\rho$-axis vector points at the methyl group, but it is not equal to the internal axis of the methyl top. This can be also seen by the comparison of the angles $\angle(i,a,b,c)$ with the torsional parameters $\rho$, $\beta$, and $\alpha$ in Table \ref{tab:resultsII}. 
\\
The inclusion of the energy tunneling parameter $\epsilon_{2}$ did not improve the fit and thus was omitted. 
The inclusion of higher order centrifugal distortion constants, like $h_{J}$ and $h_{K}$ that were used in the GS analysis \citep{Mesko.2017}, resulted in large uncertainties (about 50\%) and did not improve the fit, so they were also omitted.
\\
Concerning the error analysis of the fit, the frequency resolution $\Delta x_{obs}$ of each band was used as the error for the weighted root-mean-square (WRMS) error $\sigma_{err}$ analysis\footnote{The WRMS error is defined as $\sigma_{err}=\sqrt{\frac{1}{N}\sum \left((x_{obs}-x_{calc})/\Delta x_{obs}\right)^2}$.}. The microwave RMS\footnote{The microwave RMS is the 1$\sigma$ uncertainty given by the equation $\sigma=\sqrt{\frac{1}{N}\sum \left(x_{obs}-x_{calc}\right)^2}$.} $\sigma$ was also determined. ERHAM considers the uncertainty value of each line, but calculates the mean value if the lines are blended.
Thus, the number of lines considered in the uncertainty of the fit is lower compared to the number of lines of the XIAM analysis. 
\\
\\
The internal rotation of PO is characterised by a potential function with a threefold symmetry that can be written as \citep{LIN.1959,Lister.1978}:
\begin{eqnarray}
V(\alpha')=\sum\limits_{n=3,6,...}^{\infty}\dfrac{1}{2}V_{n}(1-\cos(n\alpha'))\approx \dfrac{V_{3}}{2}(1-\cos(3\alpha')) \label{pot}
\end{eqnarray}
where $\alpha'$ is the angle of rotation of the methyl group in comparison to the rest framework of the molecule, $n$ is the $n$-fold symmetry, and $V_{n}$ are the respective barrier heights. In the case of PO, the $V_{6}$ contribution was bound to $V_{6}<10~\mathrm{cm^{-1}}$, and the relative $V_{6}$ contribution to the A-E splitting decreases with higher excited torsional states, such as $\mathrm{\upsilon_{24}}$, which has been shown previously \citep{Herschbach.1958,Herschbach.1957,Herschbach.1957b,Herschbach.1959}. Therefore, the potential function can be approximated by considering only the $V_{3}$ contribution.  
The Schrödinger equation together with the potential function given by Equation (\ref{pot}) can be formulated as a Mathieu equation with the following correlations in the high barrier approximation \citep{LIN.1959,Lister.1978,Herschbach.1958,Herschbach.1957,Wilson.1957}:
\begin{eqnarray}
V_{n}=\dfrac{n^{2}Fs}{4}~\overset{n=3}{=}~\dfrac{9}{4}Fs   \label{barrier}
\end{eqnarray} 
\begin{eqnarray}
F=\dfrac{h^{2}}{8\pi^{2}rI_{\alpha}} 
\end{eqnarray} 
\begin{eqnarray}
r=1-\sum\limits_{i=a,b,c}^{}\frac{\lambda_{i}^{2}I_{\alpha}}{I_{i}}
\end{eqnarray} 
\begin{eqnarray}
F_{0}=\dfrac{h^{2}}{8\pi^{2}I_{\alpha}  }
\label{form:F0}
\end{eqnarray}
\begin{eqnarray}
I_{r}=rI_{\alpha}
\end{eqnarray} 
where $h$ is Planck's constant, $s$ is the reduced barrier height, $F$ is the reduced rotational constant, $I_{\alpha}$ is the internal moment of inertia of the symmetric top ($-\mathrm{CH_{3}}$), and $\lambda_{i}$ ($i=a,b,c$) are the direction cosines between the internal axis and the principal axis ($a,b,c$), which are calculated during the internal rotation analysis in ERHAM and XIAM. The reduced rotational constant $F_{0}$ was derived from the internal moment of inertia $I_{\alpha}$ and the direction cosines of $\angle(i,a/b/c)$ during the ERHAM analysis. With the ERHAM results for $F$ given in Table \ref{tab:resultsII} and the value $s=68.0$ from Herschbach \& Swalen \citep{Herschbach.1958}, the potential barrier height was calculated in the high barrier approximation\footnote{The uncertainty of the $V_{3}$ calculation depends on the uncertainty of $s=68.0$, which was not reported. Assuming an uncertainty of $\delta s=1$, the uncertainty of $V_{3}$ can be calculated to be $\delta V_{3}=$13.282$\,\mathrm{cm^{-1}}$.}, using Equation (\ref{barrier}), to be  $V_{3}=898.154\,\mathrm{cm^{-1}}$, which is close to the deduced GS values reported by Herschbach and Swalen \citep{Herschbach.1958} and Mesko \textit{et al.} \citep{Mesko.2017} ($V_{3}=895(5)\,\mathrm{cm^{-1}}$ and $V_{3}=892.71(58)\,\mathrm{cm^{-1}}$, respectively). 
\\
\\
In order to fit the potential barrier height $V_{3}$, the program XIAM was used, in which the same rotational constants up to the sextic order (except for $\Phi_{K}$) were included as for the ERHAM analysis. In addition, we included three tunneling parameters, $\Delta_{\pi2J}$, $\Delta_{\pi2K}$, and $\Delta_{\pi2-}$, which are defined by Hansen \textit{et al.} \citep{Hansen.1999}. In the XIAM analysis, the parameters $\rho_{X}$, $\beta_{X}$, $\gamma_{X}$, and $F$ are derived from the fit, and $F_{0}$ is taken from Herschbach and Swalen \citep{Herschbach.1958}, and thus no uncertainty is given. In the XIAM analysis, the reduced rotational constant $F_{0}$ was fixed to $F_{0}=158.2278$\,GHz. 
The threefold barrier height tunneling parameter $V_{3}$ was calculated to be $V_{3}=894.5079(259)\,\mathrm{cm^{-1}}$.

\section{Discussion} \label{disc}
\subsection{Quantum chemical calculations \& experiment}
The calculated rotational constants for the 1$^{st}$ ETS (Table \ref{tab:abini}) show a deviation from the experimental values (Table \ref{tab:resultsI}) in the order of 2$\%$. 
The rotational GS quartic centrifugal distortion constants in Table \ref{tab:abini} are in good agreement with the experimentally derived values of the 1$^{st}$ ETS with typically less than 5\% deviation.
The sextic centrifugal distortion constants show larger deviations of 5\% to 10\%. 
Deviations from the calculated and literature values \citep{Mesko.2017} can be explained by the fact that, generally, the GS and ETS parameters differ, and because of the influence of the respective tunneling parameters in ERHAM and XIAM on the rotational centrifugal distortion constants. In total, the MP2/def2-TZVP calculation shows the smallest deviation from experiment with regard to the rotational constants, whereas the quartic centrifugal distortion parameters from the B3LYP/aug-cc-pVTZ calculations were closest to the experimental values. Using the B3LYP/def2-TZVP level of theory, a hindered internal motion calculation was performed. The barrier height was calculated to be $V_{3}^{calc}=2.48\,\mathrm{kcal}=10.37\,\mathrm{kJ}=867.21\,\mathrm{cm^{-1}}$. The harmonically calculated value is about 30 $\mathrm{cm^{-1}}$ below the experimentally derived XIAM value $V_{3}=894.5079(259)\,\mathrm{cm^{-1}}$ of the 1$^{st}$ ETS in Table \ref{tab:resultsII}, which corresponds to a deviation of 3\,\%. 
\subsection{ERHAM analysis \& literature}
We used ERHAM to determine molecular, tunneling, and structural parameters of PO and achieved a 1$\sigma$ confidence level of 87.3\,kHz for a total of 4906 lines from 75\,GHz to 950\,GHz.
The ERHAM rotational constants are close to the values from Herschbach and Swalen \citep{Swalen.1957,Herschbach.1958}, where the A and E states were fit separately with a common set of (ground state) centrifugal distortion constants. 
The ERHAM rotational constant $A=18014.521213$ MHz is close to the arithmetic mean $\overline{A}=\left[ A_{\mathrm{A}}+2A_{\mathrm{E}}\right]/3=$18014.37\,MHz of the  rotational constant $A$ of the A and E state.
This is also true for the rotational constants $B$ and $C$. 
The reported ground state centrifugal distortion constants from Herschbach and Swalen \citep{Herschbach.1958} are close to the parameters yielded with ERHAM for the 1$^{st}$ ETS of PO. 
There is good agreement between the ERHAM results and the literature \citep{Herschbach.1958} concerning the structural parameters (Table \ref{tab:resultsII}): all parameters show deviations of less than 1\%. The structure of the internal methyl top and the $\rho$-axis vector are well defined with respect to the principal axis system, as can be seen from the uncertainty of the respective angles $\angle(i,a/b/c)$ and the $\rho$ value. The value of the torsional energy difference $\Delta_{0}$, which describes the size of the splitting of the A and E states of the respective torsional state, derived in this work and in Herschbach and Swalen \citep{Herschbach.1958} show a deviation of less than 1.2\,\%.      
\subsection{ERHAM vs. XIAM analysis}
The results of the ERHAM and XIAM analysis of the 1$^{st}$ ETS of PO, see Tables \ref{tab:resultsI} and \ref{tab:resultsII}, are based on almost the same line list in their respective input files. Thus, many results can be directly compared. 
In general, the ERHAM and XIAM results are in good agreement. The primary rotational constants as well as the centrifugal distortion constants are well determined, and most parameters match within the given uncertainty. Only a few values do not overlap within the 1$\sigma$ uncertainty range, for example the centrifugal distortion parameters: $\Delta_{K}$, $\delta_{K}$, and $\Phi_{KJ}$. In our XIAM analysis, $\Phi_{K}$ was not included as it caused large errors in other centrifugal distortion and tunneling parameters. The differences in the determined distortion parameters can be explained by the use of  different tunneling parameters in ERHAM and XIAM. In the ground state analysis \citep{Mesko.2017}, the barrier height $V_{3}$ and the tunneling parameter $\Delta_{\pi2K}$ were neccessary to achieve a satisfactory result. In our XIAM analysis of the 1$^{st}$ ETS, we needed the additional tunneling parameters $\Delta_{\pi2J}$ and $\Delta_{\pi2-}$ to satisfactorily describe this LAM. 
The inclusion of the additional internal rotation parameters $D_{c3J}$, $D_{c3K}$, and $D_{c3-}$, which were introduced and successfully applied by Herbers \textit{et al.} in the analysis of 4-methylacetophenone   \citep{Herbers.2020,Herbers.2020b}, did not improve the fit of the rotational transitions of PO's 1$^{st}$ ETS and thus were not used in this work. 
\\
Furthermore, in the XIAM analysis, we fixed $F_{0}=158.2278$\,GHz (see Herschbach \& Swalen \citep{Herschbach.1958} and Mesko \textit{et al.} \citep{Mesko.2017}), and $\rho, F, s,  \gamma_{X},$ and $\beta_{X}$ were derived from the internal rotor analysis. Beside these differences, the $\rho$ value and the angles $\angle(i,a/b/c)$ are close to the results of our ERHAM fit. The $F$ value is consistent for all studies compared in Table \ref{tab:resultsII}. 
The barrier height $V_{3}$ from our XIAM analysis matches that of Herschbach and Swalen \citep{Herschbach.1958} within the 1$\sigma$ uncertainty and within a 3$\sigma$         
uncertainty of the value reported in Mesko \textit{et al.} \citep{Mesko.2017}. The $V_{6}$ tunneling parameter was omitted since it did not improve the fit and resulted in large uncertainties. This was also discussed elsewhere  \citep{Mesko.2017,Herschbach.1958,Herschbach.1957,Herschbach.1957b,Herschbach.1959}.  
\\
The XIAM results have a slightly larger uncertainty than the respective parameters obtained from ERHAM. The XIAM uncertainty $\sigma_{X}=182.9$\,kHz is more than twice the uncertainty $\sigma_{E}=87.3$\,kHz of the ERHAM fit.
This can be explained as follows: on the one hand, in contrast to ERHAM, XIAM does not calculate the mean uncertainty of blended lines (see Section \ref{specres}), which can result in higher uncertainties in the XIAM analysis for equal assignments. On the other hand, there are less tunneling parameters available in XIAM. Specifically, when treating torsionally excited states with XIAM, the resulting large splittings are frequently less well described and therefore a fit to experimental uncertainty may not be achieved. A similar behaviour has been reported by Kisiel \textit{et al.} \citep{Kisiel.2007} in the case of pyruvic acid, where the ground state was fit with both ERHAM and XIAM, but a satisfactory fit of the excited states was only achieved using ERHAM and SPFIT \citep{Pickett.1991}. ERHAM uses more tunneling parameters than XIAM to describe the splittings, and it is well suited to analyse large splittings like those of PO's torsionally excited states. The rotational and centrifugal distortion tunneling parameters allow for a flexible correction of the rotational parameters $A$, $B$, $C$, ..., $\Phi$ caused by the effects of internal rotation. The energy tunneling parameter $\epsilon$ is also strongly linked to the size of the A-E splittings. 
With ERHAM, it is not possible to directly determine the barrier height $V_{3}$, however, this is possible with XIAM, and the joint analysis with both programs allows for a comprehensive study of the internal rotation of PO.
\\
In summary, both programs (XIAM and ERHAM) are well suited to analyse the observed spectra, as can be seen in Figures \ref{fig:100}, \ref{fig:200}, and \ref{fig:800}.

\subsection{Radioastronomical searches}
PO has been detected in space towards Sgr B2(N) \citep{McGuire.2016}; the modelling temperature $T\approx 5$\,K is rather low, and thus, the first excited torsional state is barely populated. In contrast, in warm environments with temperatures above $T=100$\,K, for example in hot core regions, PO's first vibrationally excited state, $\upsilon_{24}$, with $\nu\approx 200\,\mathrm{cm^{-1}}$ would be excited if PO is present and signals should be detectable by means of radioastronomy in this case. 
The ERHAM and XIAM results of the analysis of the first excited torsional state of PO from this work allow the assignment of the dense spectra of PO in the mmw and sub-mmw region. Our predictions could help radioastronomers with the assignment of the strong lines belonging to PO's first excited torsional state and hence could allow them to decipher its dense spectra in warm astronomical sources. Promising targets could be sources similar to Orion-KL, where PO's isomer acetone has been detected \citep{Friedel.2005}, as well as IRAS 16293-2422, where its isomers acetone and propanal, and the molecule oxirane have been detected \citep{Lykke.2017}.

\section{Conclusions}\label{conc}
We present a high-resolution spectral analysis of the chiral molecule propylene oxide with observed transitions from 10$-$950\,GHz in its 1$^{st}$ ETS. We included about 55 lines from Herschbach and Swalen \cite{Herschbach.1958} and assigned more than 4900 A and E state transitions with $(J, K_{a})<(56, 25)$ between 75 GHz and 950 GHz, resulting in an accurate prediction of the splitting of the first excited torsional state of PO up to 1 THz using ERHAM and XIAM. The anharmonic frequency calculations performed for this work and the literature values \citep{Mesko.2017,Herschbach.1958} are in good agreement with the determined rotational constants. Furthermore, we determined the tunneling paremeters using the programs ERHAM and XIAM, which describe the A-E splittings throughout the whole frequency range up to 1 THz with a microwave RMS of about 87 kHz and 183 kHz, respectively. The internal motion analysis enabled a precise geometrical description of the internal axis with respect to the principal axis frame. We determined the threefold barrier height $V_{3}=894.5709(259)~\mathrm{cm^{-1}}$ using XIAM and compared our results to literature values. 
The molecular constants are summarised in Tables \ref{tab:resultsI} and \ref{tab:resultsII} and have been discussed and compared with literature values. For the PO analysis, the ERHAM results seem to be more precise compared to the XIAM fit. However, the XIAM analysis allowed us to directly determine the barrier height $V_{3}$.
The frequency uncertainty of ERHAM and XIAM, and the overall comparison of our results, their agreement with theory and literature, including about 5000 transitions up to 1 THz, show that we have found a robust parameter set for both analysis methods.
\\
The results presented here help to better understand the dense spectrum of PO and can assist astronomers in their identification of rotational transitions originating from the first excited torsional state of this molecule in the interstellar medium.

\section{Acknowledgements}
This paper is dedicated to Stephan Schlemmer of the University of Cologne on the occasion of his 60$^{th}$ birthday. Stephan is not only an excellent scientist, but also a kind friend and colleague who is always happy to support us with advice and assistance. His enthusiasm for laboratory astrophysics is a true inspiration for young as well as experienced researchers.
\\
\\
TFG, MS, and PS acknowledge the funding by the Deutsche Forschungsgemeinschaft (DFG, German Research Foundation) – Projektnummer 328961117 – SFB 1319 ELCH.  PS and GWF also gratefully acknowledge the funding from the DFG-FU 715/3-1 project. OZ is grateful to the funding by the Collaborative Research Centre 956 funded by the Deutsche Forschungsgemeinschaft (DFG) – project ID 184018867. MS acknowledges the funding by the ERC Starting Grant ASTROROT (638027).
BEA gives thanks to the International Max Planck Research School for Ultrafast Imaging and Structural Dynamics (IMPRS-UFAST).  
LM/RAM acknowledge the support of the Programme National "Physique et Chimie du Milieu Interstellaire" (PCMI) of CNRS/INSU. 
PS acknowledges the support by Peter Groner (University of Missouri-Kansas City). Thanks to Stewart Novick from the Wesleyan University and Daniel Obenchain (formerly DESY, now Georg-August University) for their interesting talks about internal rotors and their investigation.


\begin{thebibliography}{42}
\expandafter\ifx\csname natexlab\endcsname\relax\def\natexlab#1{#1}\fi
\providecommand{\url}[1]{\texttt{#1}}
\providecommand{\href}[2]{#2}
\providecommand{\path}[1]{#1}
\providecommand{\DOIprefix}{doi:}
\providecommand{\ArXivprefix}{arXiv:}
\providecommand{\URLprefix}{URL: }
\providecommand{\Pubmedprefix}{pmid:}
\providecommand{\doi}[1]{\href{http://dx.doi.org/#1}{\path{#1}}}
\providecommand{\Pubmed}[1]{\href{pmid:#1}{\path{#1}}}
\providecommand{\bibinfo}[2]{#2}
\ifx\xfnm\relax \def\xfnm[#1]{\unskip,\space#1}\fi
\bibitem[{Fox and Jennings(1978)}]{Fox.1978}
\bibinfo{author}{K.~Fox}, \bibinfo{author}{D.~E. Jennings},
  \bibinfo{journal}{The Astrophysical Journal} \bibinfo{volume}{226}
  (\bibinfo{year}{1978}) \bibinfo{pages}{L43}.
  \DOIprefix\doi{\url{10.1086/182827}}.
\bibitem[{Dickens et~al.(1997)Dickens, Irvine, Ohishi, Ikeda, Ishikawa,
  Nummelin, and Hjalmarson}]{Dickens.1997}
\bibinfo{author}{J.~E. Dickens}, \bibinfo{author}{W.~M. Irvine},
  \bibinfo{author}{M.~Ohishi}, \bibinfo{author}{M.~Ikeda},
  \bibinfo{author}{S.~Ishikawa}, \bibinfo{author}{A.~Nummelin},
  \bibinfo{author}{A.~Hjalmarson}, \bibinfo{journal}{The Astrophysical Journal}
  \bibinfo{volume}{489} (\bibinfo{year}{1997}) \bibinfo{pages}{753--757}.
  \DOIprefix\doi{\url{10.1086/304821}}.
\bibitem[{McGuire et~al.(2016)McGuire, Carroll, Loomis, Finneran, Jewell,
  Remijan, and Blake}]{McGuire.2016}
\bibinfo{author}{B.~A. McGuire}, \bibinfo{author}{P.~B. Carroll},
  \bibinfo{author}{R.~A. Loomis}, \bibinfo{author}{I.~A. Finneran},
  \bibinfo{author}{P.~R. Jewell}, \bibinfo{author}{A.~J. Remijan},
  \bibinfo{author}{G.~A. Blake}, \bibinfo{journal}{Science (New York, N.Y.)}
  \bibinfo{volume}{352} (\bibinfo{year}{2016}) \bibinfo{pages}{1449--1452}.
  \DOIprefix\doi{\url{10.1126/science.aae0328}}.
\bibitem[{Herbst and {van Dishoeck}(2009)}]{Herbst.2009}
\bibinfo{author}{E.~Herbst}, \bibinfo{author}{E.~F. {van Dishoeck}},
  \bibinfo{journal}{Annual Review of Astronomy and Astrophysics}
  \bibinfo{volume}{47} (\bibinfo{year}{2009}) \bibinfo{pages}{427--480}.
  \DOIprefix\doi{\url{10.1146/annurev-astro-082708-101654}}.
\bibitem[{Snyder et~al.(2002)Snyder, Lovas, Mehringer, Miao, Kuan, Hollis, and
  Jewell}]{Snyder.2002}
\bibinfo{author}{L.~E. Snyder}, \bibinfo{author}{F.~J. Lovas},
  \bibinfo{author}{D.~M. Mehringer}, \bibinfo{author}{N.~Y. Miao},
  \bibinfo{author}{Y.-J. Kuan}, \bibinfo{author}{J.~M. Hollis},
  \bibinfo{author}{P.~R. Jewell}, \bibinfo{journal}{The Astrophysical Journal}
  \bibinfo{volume}{578} (\bibinfo{year}{2002}) \bibinfo{pages}{245--255}.
  \DOIprefix\doi{\url{10.1086/342273}}.
\bibitem[{Friedel et~al.(2005)Friedel, Snyder, Remijan, and
  Turner}]{Friedel.2005}
\bibinfo{author}{D.~N. Friedel}, \bibinfo{author}{L.~E. Snyder},
  \bibinfo{author}{A.~J. Remijan}, \bibinfo{author}{B.~E. Turner},
  \bibinfo{journal}{The Astrophysical Journal} \bibinfo{volume}{632}
  (\bibinfo{year}{2005}) \bibinfo{pages}{L95--L98}.
  \DOIprefix\doi{\url{10.1086/497986}}.
\bibitem[{Lykke et~al.(2017)Lykke, Coutens, J{\o}rgensen, {van der Wiel},
  Garrod, M{\"u}ller, Bjerkeli, Bourke, Calcutt, Drozdovskaya, Favre, Fayolle,
  Jacobsen, {\"O}berg, Persson, {van Dishoeck}, and Wampfler}]{Lykke.2017}
\bibinfo{author}{J.~M. Lykke}, \bibinfo{author}{A.~Coutens},
  \bibinfo{author}{J.~K. J{\o}rgensen}, \bibinfo{author}{M.~H.~D. {van der
  Wiel}}, \bibinfo{author}{R.~T. Garrod}, \bibinfo{author}{H.~S.~P.
  M{\"u}ller}, \bibinfo{author}{P.~Bjerkeli}, \bibinfo{author}{T.~L. Bourke},
  \bibinfo{author}{H.~Calcutt}, \bibinfo{author}{M.~N. Drozdovskaya},
  \bibinfo{author}{C.~Favre}, \bibinfo{author}{E.~C. Fayolle},
  \bibinfo{author}{S.~K. Jacobsen}, \bibinfo{author}{K.~I. {\"O}berg},
  \bibinfo{author}{M.~V. Persson}, \bibinfo{author}{E.~F. {van Dishoeck}},
  \bibinfo{author}{S.~F. Wampfler}, \bibinfo{journal}{Astronomy {\&}
  Astrophysics} \bibinfo{volume}{597} (\bibinfo{year}{2017})
  \bibinfo{pages}{A53}. \DOIprefix\doi{\url{10.1051/0004-6361/201629180}}.
\bibitem[{Mesko et~al.(2017)Mesko, Zou, Carroll, and {Widicus
  Weaver}}]{Mesko.2017}
\bibinfo{author}{A.~J. Mesko}, \bibinfo{author}{L.~Zou}, \bibinfo{author}{P.~B.
  Carroll}, \bibinfo{author}{S.~L. {Widicus Weaver}}, \bibinfo{journal}{Journal
  of Molecular Spectroscopy} \bibinfo{volume}{335} (\bibinfo{year}{2017})
  \bibinfo{pages}{49--53}. \DOIprefix\doi{\url{10.1016/j.jms.2017.02.003}}.
\bibitem[{Hartwig and Dreizler(1996)}]{Hartwig.1996}
\bibinfo{author}{H.~Hartwig}, \bibinfo{author}{H.~Dreizler},
  \bibinfo{journal}{Zeitschrift f{\"u}r Naturforschung A} \bibinfo{volume}{51}
  (\bibinfo{year}{1996}). \DOIprefix\doi{\url{10.1515/zna-1996-0807}}.
\bibitem[{LIN and Swalen(1959)}]{LIN.1959}
\bibinfo{author}{C.~C. LIN}, \bibinfo{author}{J.~D. Swalen},
  \bibinfo{journal}{Reviews of Modern Physics} \bibinfo{volume}{31}
  (\bibinfo{year}{1959}) \bibinfo{pages}{841--892}.
  \DOIprefix\doi{\url{10.1103/RevModPhys.31.841}}.
\bibitem[{Lister et~al.(1978)Lister, Macdonald, and Owen}]{Lister.1978}
\bibinfo{author}{D.~G. Lister}, \bibinfo{author}{J.~N. Macdonald},
  \bibinfo{author}{N.~L. Owen}, \bibinfo{title}{Internal rotation and
  inversion: An introd. to large amplitude motions in molecules},
  \bibinfo{publisher}{{Acad. Pr}}, \bibinfo{address}{London},
  \bibinfo{year}{1978}.
\bibitem[{Swalen and Herschbach(1957)}]{Swalen.1957}
\bibinfo{author}{J.~D. Swalen}, \bibinfo{author}{D.~R. Herschbach},
  \bibinfo{journal}{The Journal of Chemical Physics} \bibinfo{volume}{27}
  (\bibinfo{year}{1957}) \bibinfo{pages}{100--108}.
  \DOIprefix\doi{\url{10.1063/1.1743645}}.
\bibitem[{Herschbach and Swalen(1958)}]{Herschbach.1958}
\bibinfo{author}{D.~R. Herschbach}, \bibinfo{author}{J.~D. Swalen},
  \bibinfo{journal}{The Journal of Chemical Physics} \bibinfo{volume}{29}
  (\bibinfo{year}{1958}) \bibinfo{pages}{761--776}.
  \DOIprefix\doi{\url{10.1063/1.1744588}}.
\bibitem[{Creswell and Schwendeman(1977)}]{Creswell.1977}
\bibinfo{author}{R.~A. Creswell}, \bibinfo{author}{R.~H. Schwendeman},
  \bibinfo{journal}{Journal of Molecular Spectroscopy} \bibinfo{volume}{64}
  (\bibinfo{year}{1977}) \bibinfo{pages}{295--301}.
  \DOIprefix\doi{\url{10.1016/0022-2852(77)90268-5}}.
\bibitem[{Imachi and Kuczkowski(1983)}]{Imachi.1983}
\bibinfo{author}{M.~Imachi}, \bibinfo{author}{R.~L. Kuczkowski},
  \bibinfo{journal}{Journal of Molecular Structure} \bibinfo{volume}{96}
  (\bibinfo{year}{1983}) \bibinfo{pages}{55--60}.
  \DOIprefix\doi{\url{10.1016/0022-2860(82)90057-6}}.
\bibitem[{Groner(1997)}]{Groner.1997}
\bibinfo{author}{P.~Groner}, \bibinfo{journal}{The Journal of Chemical Physics}
  \bibinfo{volume}{107} (\bibinfo{year}{1997}) \bibinfo{pages}{4483--4498}.
  \DOIprefix\doi{\url{10.1063/1.474810}}.
\bibitem[{Groner(2012)}]{Groner.2012}
\bibinfo{author}{P.~Groner}, \bibinfo{journal}{Journal of Molecular
  Spectroscopy} \bibinfo{volume}{278} (\bibinfo{year}{2012})
  \bibinfo{pages}{52--67}. \DOIprefix\doi{\url{10.1016/j.jms.2012.06.006}}.
\bibitem[{Arenas et~al.(2017)Arenas, Gruet, Steber, Giuliano, and
  Schnell}]{Arenas.2017}
\bibinfo{author}{B.~E. Arenas}, \bibinfo{author}{S.~Gruet},
  \bibinfo{author}{A.~L. Steber}, \bibinfo{author}{B.~M. Giuliano},
  \bibinfo{author}{M.~Schnell}, \bibinfo{journal}{Phys. Chem. Chem. Phys.}
  \bibinfo{volume}{19} (\bibinfo{year}{2017}) \bibinfo{pages}{1751--1756}.
  \DOIprefix\doi{\url{10.1039/C6CP06297K}}.
\bibitem[{Neill et~al.(2013)Neill, Harris, Steber, Douglass, Plusquellic, and
  Pate}]{Neill.2013}
\bibinfo{author}{J.~L. Neill}, \bibinfo{author}{B.~J. Harris},
  \bibinfo{author}{A.~L. Steber}, \bibinfo{author}{K.~O. Douglass},
  \bibinfo{author}{D.~F. Plusquellic}, \bibinfo{author}{B.~H. Pate},
  \bibinfo{journal}{Opt. Express} \bibinfo{volume}{21} (\bibinfo{year}{2013})
  \bibinfo{pages}{19743--19749}. \DOIprefix\doi{\url{10.1364/OE.21.019743}}.
\bibitem[{Bott and Sadler(1966)}]{Bott.1966}
\bibinfo{author}{T.~R. Bott}, \bibinfo{author}{H.~N. Sadler},
  \bibinfo{journal}{Journal of Chemical {\&} Engineering Data}
  \bibinfo{volume}{11} (\bibinfo{year}{1966}) \bibinfo{pages}{25}.
  \DOIprefix\doi{\url{10.1021/je60028a005}}.
\bibitem[{Herberth et~al.(2019)Herberth, Giesen, and Yamada}]{Herberth.2019}
\bibinfo{author}{D.~Herberth}, \bibinfo{author}{T.~F. Giesen},
  \bibinfo{author}{K.~Yamada}, \bibinfo{journal}{Journal of Molecular
  Spectroscopy} \bibinfo{volume}{362} (\bibinfo{year}{2019})
  \bibinfo{pages}{37--44}. \DOIprefix\doi{\url{10.1016/j.jms.2019.05.011}}.
\bibitem[{Stahl et~al.(2020)Stahl, Arenas, Domingos, Fuchs, Schnell, and
  Giesen}]{Stahl.2020}
\bibinfo{author}{P.~Stahl}, \bibinfo{author}{B.~E. Arenas},
  \bibinfo{author}{S.~R. Domingos}, \bibinfo{author}{G.~W. Fuchs},
  \bibinfo{author}{M.~Schnell}, \bibinfo{author}{T.~F. Giesen},
  \bibinfo{journal}{Physical chemistry chemical physics : PCCP}
  \bibinfo{volume}{22} (\bibinfo{year}{2020}) \bibinfo{pages}{21474--21487}.
  \DOIprefix\doi{\url{10.1039/d0cp03523h}}.
\bibitem[{Zou et~al.(2020)Zou, Motiyenko, Margul{\`e}s, and
  Alekseev}]{Zou.2020}
\bibinfo{author}{L.~Zou}, \bibinfo{author}{R.~A. Motiyenko},
  \bibinfo{author}{L.~Margul{\`e}s}, \bibinfo{author}{E.~A. Alekseev},
  \bibinfo{journal}{Review of Scientific Instruments} \bibinfo{volume}{91}
  (\bibinfo{year}{2020}) \bibinfo{pages}{063104}.
  \DOIprefix\doi{\url{10.1063/5.0004461}}.
\bibitem[{{R. A. Motiyenko} et~al.(2019){R. A. Motiyenko}, {I. A. Armieieva},
  {L. Margul{\`e}s}, {E. A. Alekseev}, and {J.-C.
  Guillemin}}]{R.A.Motiyenko.2019}
\bibinfo{author}{{R. A. Motiyenko}}, \bibinfo{author}{{I. A. Armieieva}},
  \bibinfo{author}{{L. Margul{\`e}s}}, \bibinfo{author}{{E. A. Alekseev}},
  \bibinfo{author}{{J.-C. Guillemin}}, \bibinfo{journal}{Astronomy {\&}
  Astrophysics} \bibinfo{volume}{623} (\bibinfo{year}{2019})
  \bibinfo{pages}{A162}. \URLprefix
  \url{\url{https://www.aanda.org/articles/aa/full_html/2019/03/aa34587-18/aa34587-18.html}}.
  \DOIprefix\doi{\url{10.1051/0004-6361/201834587}}.
\bibitem[{Frisch et~al.(2016)Frisch, Trucks, Schlegel, {G. E. Scuseria}, {M. A.
  Robb}, {J. R. Cheeseman}, {G. Scalmani}, {V. Barone}, {G. A. Petersson}, {H.
  Nakatsuji}, {X. Li}, {M. Caricato}, {A. V. Marenich}, {J. Bloino}, {B. G.
  Janesko}, {R. Gomperts}, {B. Mennucci}, {H. P. Hratchian}, {J. V. Ortiz}, {A.
  F. Izmaylov}, {J. L. Sonnenberg}, {D. Williams-Young}, {F. Ding}, {F.
  Lipparini}, {F. Egidi}, {J. Goings}, {B. Peng}, {A. Petrone}, {T. Henderson},
  {D. Ranasinghe}, {V. G. Zakrzewski}, {J. Gao}, {N. Rega}, {G. Zheng}, {W.
  Liang}, {M. Hada}, {M. Ehara}, {K. Toyota}, {R. Fukuda}, {J. Hasegawa}, {M.
  Ishida}, {T. Nakajima}, {Y. Honda}, {O. Kitao}, {H. Nakai}, {T. Vreven}, {K.
  Throssell}, {J. A. Montgomery}, Jr., {J. E. Peralta}, {F. Ogliaro}, {M. J.
  Bearpark}, {J. J. Heyd}, {E. N. Brothers}, {K. N. Kudin}, {V. N. Staroverov},
  {T. A. Keith}, {R. Kobayashi}, {J. Normand}, {K. Raghavachari}, {A. P.
  Rendell}, {J. C. Burant}, {S. S. Iyengar}, {J. Tomasi}, {M. Cossi}, {J. M.
  Millam}, {M. Klene}, {C. Adamo}, {R. Cammi}, {J. W. Ochterski}, {R. L.
  Martin}, {K. Morokuma}, {O. Farkas}, {J. B. Foresman}, and {D. J.
  Fox}}]{Frisch.2016}
\bibinfo{author}{M.~J. Frisch}, \bibinfo{author}{G.~W. Trucks},
  \bibinfo{author}{H.~B. Schlegel}, \bibinfo{author}{{G. E. Scuseria}},
  \bibinfo{author}{{M. A. Robb}}, \bibinfo{author}{{J. R. Cheeseman}},
  \bibinfo{author}{{G. Scalmani}}, \bibinfo{author}{{V. Barone}},
  \bibinfo{author}{{G. A. Petersson}}, \bibinfo{author}{{H. Nakatsuji}},
  \bibinfo{author}{{X. Li}}, \bibinfo{author}{{M. Caricato}},
  \bibinfo{author}{{A. V. Marenich}}, \bibinfo{author}{{J. Bloino}},
  \bibinfo{author}{{B. G. Janesko}}, \bibinfo{author}{{R. Gomperts}},
  \bibinfo{author}{{B. Mennucci}}, \bibinfo{author}{{H. P. Hratchian}},
  \bibinfo{author}{{J. V. Ortiz}}, \bibinfo{author}{{A. F. Izmaylov}},
  \bibinfo{author}{{J. L. Sonnenberg}}, \bibinfo{author}{{D. Williams-Young}},
  \bibinfo{author}{{F. Ding}}, \bibinfo{author}{{F. Lipparini}},
  \bibinfo{author}{{F. Egidi}}, \bibinfo{author}{{J. Goings}},
  \bibinfo{author}{{B. Peng}}, \bibinfo{author}{{A. Petrone}},
  \bibinfo{author}{{T. Henderson}}, \bibinfo{author}{{D. Ranasinghe}},
  \bibinfo{author}{{V. G. Zakrzewski}}, \bibinfo{author}{{J. Gao}},
  \bibinfo{author}{{N. Rega}}, \bibinfo{author}{{G. Zheng}},
  \bibinfo{author}{{W. Liang}}, \bibinfo{author}{{M. Hada}},
  \bibinfo{author}{{M. Ehara}}, \bibinfo{author}{{K. Toyota}},
  \bibinfo{author}{{R. Fukuda}}, \bibinfo{author}{{J. Hasegawa}},
  \bibinfo{author}{{M. Ishida}}, \bibinfo{author}{{T. Nakajima}},
  \bibinfo{author}{{Y. Honda}}, \bibinfo{author}{{O. Kitao}},
  \bibinfo{author}{{H. Nakai}}, \bibinfo{author}{{T. Vreven}},
  \bibinfo{author}{{K. Throssell}}, \bibinfo{author}{{J. A. Montgomery}},
  \bibinfo{author}{Jr.}, \bibinfo{author}{{J. E. Peralta}},
  \bibinfo{author}{{F. Ogliaro}}, \bibinfo{author}{{M. J. Bearpark}},
  \bibinfo{author}{{J. J. Heyd}}, \bibinfo{author}{{E. N. Brothers}},
  \bibinfo{author}{{K. N. Kudin}}, \bibinfo{author}{{V. N. Staroverov}},
  \bibinfo{author}{{T. A. Keith}}, \bibinfo{author}{{R. Kobayashi}},
  \bibinfo{author}{{J. Normand}}, \bibinfo{author}{{K. Raghavachari}},
  \bibinfo{author}{{A. P. Rendell}}, \bibinfo{author}{{J. C. Burant}},
  \bibinfo{author}{{S. S. Iyengar}}, \bibinfo{author}{{J. Tomasi}},
  \bibinfo{author}{{M. Cossi}}, \bibinfo{author}{{J. M. Millam}},
  \bibinfo{author}{{M. Klene}}, \bibinfo{author}{{C. Adamo}},
  \bibinfo{author}{{R. Cammi}}, \bibinfo{author}{{J. W. Ochterski}},
  \bibinfo{author}{{R. L. Martin}}, \bibinfo{author}{{K. Morokuma}},
  \bibinfo{author}{{O. Farkas}}, \bibinfo{author}{{J. B. Foresman}},
  \bibinfo{author}{{D. J. Fox}}, \bibinfo{title}{Gaussian 16},
  \bibinfo{year}{2016}.
\bibitem[{Western(2014)}]{Western.2014}
\bibinfo{author}{C.~Western}, \bibinfo{title}{Pgopher version 8.0},
  \bibinfo{year}{2014}.
  \DOIprefix\doi{\url{10.5523/BRIS.HUFLGGVPCUC1ZVLIQED497R2}}.
\bibitem[{Western(2017)}]{Western.2017}
\bibinfo{author}{C.~M. Western}, \bibinfo{journal}{Journal of Quantitative
  Spectroscopy and Radiative Transfer} \bibinfo{volume}{186}
  (\bibinfo{year}{2017}) \bibinfo{pages}{221--242}.
  \DOIprefix\doi{\url{10.1016/j.jqsrt.2016.04.010}}.
\bibitem[{Kisiel et~al.(2005)Kisiel, Pszcz{\'o}{\l}kowski, Medvedev,
  Winnewisser, de~Lucia, and Herbst}]{Kisiel.2005}
\bibinfo{author}{Z.~Kisiel}, \bibinfo{author}{L.~Pszcz{\'o}{\l}kowski},
  \bibinfo{author}{I.~R. Medvedev}, \bibinfo{author}{M.~Winnewisser},
  \bibinfo{author}{F.~C. de~Lucia}, \bibinfo{author}{E.~Herbst},
  \bibinfo{journal}{Journal of Molecular Spectroscopy} \bibinfo{volume}{233}
  (\bibinfo{year}{2005}) \bibinfo{pages}{231--243}.
  \DOIprefix\doi{\url{10.1016/j.jms.2005.07.006}}.
\bibitem[{Kisiel et~al.(2012)Kisiel, Pszcz{\'o}{\l}kowski, Drouin, Brauer, Yu,
  Pearson, Medvedev, Fortman, and Neese}]{Kisiel.2012}
\bibinfo{author}{Z.~Kisiel}, \bibinfo{author}{L.~Pszcz{\'o}{\l}kowski},
  \bibinfo{author}{B.~J. Drouin}, \bibinfo{author}{C.~S. Brauer},
  \bibinfo{author}{S.~Yu}, \bibinfo{author}{J.~C. Pearson},
  \bibinfo{author}{I.~R. Medvedev}, \bibinfo{author}{S.~Fortman},
  \bibinfo{author}{C.~Neese}, \bibinfo{journal}{Journal of Molecular
  Spectroscopy} \bibinfo{volume}{280} (\bibinfo{year}{2012})
  \bibinfo{pages}{134--144}. \DOIprefix\doi{\url{10.1016/j.jms.2012.06.013}}.
\bibitem[{Demaison et~al.(2001)Demaison, Sarka, and Cohen}]{Demaison.2001}
\bibinfo{author}{J.~Demaison}, \bibinfo{author}{K.~Sarka},
  \bibinfo{author}{E.~A. Cohen}, \bibinfo{title}{Spectroscopy from Space},
  volume~\bibinfo{volume}{20} of \textit{\bibinfo{series}{NATO Science Series,
  Series II}}, \bibinfo{publisher}{Springer}, \bibinfo{address}{Dordrecht},
  \bibinfo{year}{2001}. \DOIprefix\doi{\url{10.1007/978-94-010-0832-7}}.
\bibitem[{Pickett et~al.(1998)Pickett, Poynter, Cohen, Delitsky, Pearson, and
  M{\"u}ller}]{PICKETT.1998}
\bibinfo{author}{H.~M. Pickett}, \bibinfo{author}{R.~L. Poynter},
  \bibinfo{author}{E.~A. Cohen}, \bibinfo{author}{M.~L. Delitsky},
  \bibinfo{author}{J.~C. Pearson}, \bibinfo{author}{H.~M{\"u}ller},
  \bibinfo{journal}{Journal of Quantitative Spectroscopy and Radiative
  Transfer} \bibinfo{volume}{60} (\bibinfo{year}{1998})
  \bibinfo{pages}{883--890}.
  \DOIprefix\doi{\url{10.1016/S0022-4073(98)00091-0}}.
\bibitem[{M{\"u}ller et~al.(2005)M{\"u}ller, Schl{\"o}der, Stutzki, and
  Winnewisser}]{Muller.2005}
\bibinfo{author}{H.~S. M{\"u}ller}, \bibinfo{author}{F.~Schl{\"o}der},
  \bibinfo{author}{J.~Stutzki}, \bibinfo{author}{G.~Winnewisser},
  \bibinfo{journal}{Journal of Molecular Structure} \bibinfo{volume}{742}
  (\bibinfo{year}{2005}) \bibinfo{pages}{215--227}.
  \DOIprefix\doi{\url{10.1016/j.molstruc.2005.01.027}}.
\bibitem[{Kleiner(2010)}]{Kleiner.2010}
\bibinfo{author}{I.~Kleiner}, \bibinfo{journal}{Journal of Molecular
  Spectroscopy} \bibinfo{volume}{260} (\bibinfo{year}{2010})
  \bibinfo{pages}{1--18}. \DOIprefix\doi{\url{10.1016/j.jms.2009.12.011}}.
\bibitem[{Herschbach(1957{\natexlab{a}})}]{Herschbach.1957}
\bibinfo{author}{D.~R. Herschbach}, \bibinfo{journal}{The Journal of Chemical
  Physics} \bibinfo{volume}{27} (\bibinfo{year}{1957}{\natexlab{a}})
  \bibinfo{pages}{1420--1421}. \DOIprefix\doi{\url{10.1063/1.1744025}}.
\bibitem[{Herschbach(1957{\natexlab{b}})}]{Herschbach.1957b}
\bibinfo{author}{D.~R. Herschbach}, \bibinfo{journal}{The Journal of Chemical
  Physics} \bibinfo{volume}{27} (\bibinfo{year}{1957}{\natexlab{b}})
  \bibinfo{pages}{975}. \DOIprefix\doi{\url{10.1063/1.1743897}}.
\bibitem[{Herschbach(1959)}]{Herschbach.1959}
\bibinfo{author}{D.~R. Herschbach}, \bibinfo{journal}{The Journal of Chemical
  Physics} \bibinfo{volume}{31} (\bibinfo{year}{1959})
  \bibinfo{pages}{91--108}. \DOIprefix\doi{\url{10.1063/1.1730343}}.
\bibitem[{Wilson(1957)}]{Wilson.1957}
\bibinfo{author}{E.~B. Wilson}, \bibinfo{journal}{Proceedings of the National
  Academy of Sciences of the United States of America} \bibinfo{volume}{43}
  (\bibinfo{year}{1957}) \bibinfo{pages}{816--820}.
  \DOIprefix\doi{\url{10.1073/pnas.43.9.816}}.
\bibitem[{Hansen et~al.(1999)Hansen, M{\"a}der, and Bruhn}]{Hansen.1999}
\bibinfo{author}{N.~Hansen}, \bibinfo{author}{H.~M{\"a}der},
  \bibinfo{author}{T.~Bruhn}, \bibinfo{journal}{Molecular Physics}
  \bibinfo{volume}{97} (\bibinfo{year}{1999}) \bibinfo{pages}{587--595}.
  \DOIprefix\doi{\url{10.1080/00268979909482857}}.
\bibitem[{Herbers et~al.(2020)Herbers, Fritz, Mishra, Nguyen, and
  Zwier}]{Herbers.2020}
\bibinfo{author}{S.~Herbers}, \bibinfo{author}{S.~M. Fritz},
  \bibinfo{author}{P.~Mishra}, \bibinfo{author}{H.~V.~L. Nguyen},
  \bibinfo{author}{T.~S. Zwier}, \bibinfo{journal}{The Journal of Chemical
  Physics} \bibinfo{volume}{152} (\bibinfo{year}{2020})
  \bibinfo{pages}{074301}. \DOIprefix\doi{\url{10.1063/1.5142401}}.
\bibitem[{Herbers and Nguyen(2020)}]{Herbers.2020b}
\bibinfo{author}{S.~Herbers}, \bibinfo{author}{H.~V.~L. Nguyen},
  \bibinfo{journal}{Journal of Molecular Spectroscopy} \bibinfo{volume}{370}
  (\bibinfo{year}{2020}) \bibinfo{pages}{111289}. \URLprefix
  \url{\url{http://www.sciencedirect.com/science/article/pii/S0022285220300576}}.
  \DOIprefix\doi{\url{10.1016/j.jms.2020.111289}}.
\bibitem[{Kisiel et~al.(2007)Kisiel, Pszcz{\'o}{\l}kowski,
  Bia{\l}kowska-Jaworska, and Charnley}]{Kisiel.2007}
\bibinfo{author}{Z.~Kisiel}, \bibinfo{author}{L.~Pszcz{\'o}{\l}kowski},
  \bibinfo{author}{E.~Bia{\l}kowska-Jaworska}, \bibinfo{author}{S.~B.
  Charnley}, \bibinfo{journal}{Journal of Molecular Spectroscopy}
  \bibinfo{volume}{241} (\bibinfo{year}{2007}) \bibinfo{pages}{220--229}.
  \DOIprefix\doi{\url{10.1016/j.jms.2006.12.011}}.
\bibitem[{Pickett(1991)}]{Pickett.1991}
\bibinfo{author}{H.~M. Pickett}, \bibinfo{journal}{Journal of Molecular
  Spectroscopy} \bibinfo{volume}{148} (\bibinfo{year}{1991})
  \bibinfo{pages}{371--377}.
  \DOIprefix\doi{\url{10.1016/0022-2852(91)90393-O}}.

\end{thebibliography}

\end{document}